\documentclass[structabstract]{aa}  
\usepackage{graphicx}
\usepackage{lscape}
\usepackage{natbib}
\bibpunct{(}{)}{;}{a}{}{,}
\usepackage{txfonts}
\usepackage{color}
\titlerunning{VLT/ISAAC near-infrared spectroscopy of YSOs in the LMC}
\authorrunning{Shimonishi et al.}

\begin{document} 

   \title{VLT/ISAAC infrared spectroscopy of embedded high-mass YSOs in the Large Magellanic Cloud: Methanol and the 3.47 $\mu$m band
          \thanks{Based on observations with the Very Large Telescope at the European Southern Observatory, Paranal, Chile (ESO programme ID 090.C-0497)}
          }

%   \subtitle{}

   \author{T. Shimonishi
             \inst{1,}\inst{2}
             \and
             E. Dartois
             \inst{3}
             \and
             T. Onaka
             \inst{4}
             \and
             F. Boulanger
             \inst{3}
             }

   \institute{Frontier Research Institute for Interdisciplinary Sciences, Tohoku University, Aramakiazaaoba 6-3, Aoba-ku, Sendai, Miyagi, 980-8578, Japan
   \and 
                Astronomical Institute, Tohoku University, Aramakiazaaoba 6-3, Aoba-ku, Sendai, Miyagi, 980-8578, Japan, 
                \email{shimonishi@astr.tohoku.ac.jp}
   \and 
               Institut d'Astrophysique Spatiale, UMR 8617, Universit\'{e} Paris-Sud, b\^{a}timent 121, 91405 Orsay, France
   \and 
               Department of Astronomy, Graduate School of Science, The University of Tokyo, 7-3-1 Hongo, Bunkyo-ku, Tokyo 113-0033, Japan
             }

   \date{}

% \abstract{}{}{}{}{} 
% 5 {} token are mandatory
 
  \abstract
   {}
   {This study aims to elucidate a possible link between chemical properties of ices in star-forming regions and environmental characteristics (particularly metallicity) of the host galaxy. 
The Large Magellanic Cloud (LMC) is an excellent target to study properties of interstellar and circumstellar medium in a different galactic environment thanks to its proximity and low metallicity. 
   }
   {We performed near-infrared, L-band spectroscopic observations toward embedded high-mass young stellar objects (YSOs) in the LMC with the Infrared Spectrometer And Array Camera (ISAAC) at the Very Large Telescope. 
The 3.2--3.7 $\mu$m spectral region, which is accessible from ground-based telescopes, is important for ice studies, since various C--H stretching vibrations of carbon bearing species fall in this region. 
   }
   {We obtained medium-resolution ($R$ $\sim$500) spectra in the 3--4 $\mu$m range for nine high-mass YSOs in the LMC. 
Additionally, we analyzed archival ISAAC data of two LMC YSOs. 
We detected absorption bands due to solid H$_2$O and CH$_3$OH as well as the 3.47 $\mu$m absorption band. 
The properties of these bands are investigated based on comparisons with Galactic embedded sources. 
The 3.53 $\mu$m CH$_3$OH ice absorption band for the LMC YSOs is found to be absent or very weak compared to that seen toward Galactic sources. 
The absorption band is weakly detected for two out of eleven objects. 
We estimate the abundance of the CH$_3$OH ice, which suggests that solid CH$_3$OH is less abundant for high-mass YSOs in the LMC than those in our Galaxy. 
The 3.47 $\mu$m absorption band is detected toward six out of eleven LMC YSOs. 
We found that the 3.47 $\mu$m band and the H$_2$O ice band correlate similarly between the LMC and Galactic samples, but the LMC sources seem to require a slightly higher H$_2$O ice threshold for the appearance of the 3.47 $\mu$m band. 
For the LMC sources with relatively large H$_2$O ice optical depths, we found that the strength ratio of the 3.47 $\mu$m band relative to the water ice band is only marginally lower than those of the Galactic sources. 
   }
   {We propose that grain surface reactions at a relatively high dust temperature (warm ice chemistry) are responsible for the observed characteristics of ice chemical compositions in the LMC; i.e., the low abundance of solid CH$_3$OH presented in this work as well as the high abundance of solid CO$_2$ reported in previous studies. 
We suggest that this warm ice chemistry is one of the important characteristics of interstellar and circumstellar chemistry in low metallicity environments. 
The low abundance of CH$_3$OH in the solid phase implies that formation of complex organic molecules from methanol-derived species is less efficient in the LMC. 
For the 3.47 $\mu$m band, the observed difference in the water ice threshold may suggest that a more shielded environment is necessary for the formation of the 3.47 $\mu$m band carrier in the LMC. 
On the one hand, in well-shielded regions of the LMC, our results suggest that the lower metallicity and different interstellar environment of the LMC have little effect on the abundance ratio of the 3.47 $\mu$m band carrier and water ice. 
   }

   \keywords{Astrochemistry -- circumstellar matter -- ISM: abundances -- ISM: molecules -- Magellanic Clouds -- Infrared: ISM}

   \maketitle
%
%________________________________________________________________

\section{Introduction}
Understanding the chemical diversity of materials in star- and planet-forming regions is one of the key topics of present-day astronomy. 
Formation of stars and planets can occur in various kinds of galaxies, which differ in size, shape, age, and environment. 
It is thus important to understand how galactic characteristics affect properties of interstellar and circumstellar medium.

It is known that infrared spectra of embedded young stellar objects (YSOs) show absorption bands due to ices in which a large amount of heavy elements and complex molecules are preserved \citep[e.g., ][]{vDB98,Boo04}. 
A variety of ice species (H$_2$O, CO$_2$, CO, CH$_3$OH, CH$_4$) have been detected toward molecular clouds and YSOs \citep[e.g.,][]{Gib04,Dar05,Pon08,Boo08,Boo11,Obe11}. 
These kinds of molecular species are also detected toward solar system objects \citep[e.g., comets,][]{Oot12}, and  ices are believed to deliver important volatile molecules to planets \citep{Ehr00b}. 
Furthermore, ices are also detected toward the central region of nearby external galaxies \citep[e.g.,][]{Spo03,Yam11,Yam13}. 

Ice chemistry plays an essential role in the overall chemical evolution of molecular clouds since it efficiently proceeds in dense and cold regions (n$_H$ $\geq$ 10$^4$ cm$^{-3}$, T $\sim$10K). 
Grain surface reactions as well as radiolysis and photolysis can produce complex molecules that are different from those produced by gas-phase reactions \citep[e.g.,][]{Tie82,Has92}. 
Therefore, the chemical properties of ices should not be neglected  to understand diversities of materials found in star- and planet-forming regions.

Observations of ices around extragalactic YSOs aim to understand which environment parameters of galaxies are relevant to ice chemistry. 
It is highly probable that different galactic environments (e.g., metallicity, radiation field) could affect the properties of circumstellar materials because circumstellar materials are closely related to interstellar medium. 
In particular, it is important to understand the impact of low metallicity environments on the interstellar and circumstellar chemistry. 
These studies are essential to constrain chemical processes occurring in the past universe, since cosmic metallicity is believed to be increasing with the evolution of the universe \citep[e.g., ][]{Raf12}. 

The Large Magellanic Cloud (LMC) is the nearest external star-forming galaxy \citep[d = 49.97 $\pm$ 1.11 kpc, ][]{Pie13} and a prime target for this study. 
The LMC has been an excellent target for studies of molecular cloud evolution and star formation \citep[e.g.,][]{Mei06,FK10}. 
It is well known that the metallicity of the interstellar medium (ISM) within the LMC is about half of the solar neighborhood \citep [e.g.,][]{Duf82,Wes90,And01,Rol02}. 
Far-infrared and sub-millimeter observations show that dust temperatures in the LMC are, on average, higher than those in our Galaxy \citep[e.g., ][]{Agu03,Sak06}. 
Further, gamma-ray observations of the LMC indicate that the cosmic-ray density in the LMC is 20$\%$ to 30$\%$ lower than typical Galactic values \citep{Abd10}. 

Previous studies have reported properties of ices around embedded high-mass YSOs in the LMC \citep{ST,ST10,ST12,ST13,thesis,vanL05,vanL10, Oli06,Oli09,Oli11,Sea09,Sea10}. 
\citet{ST10}, for example, report the results of 2.5--5.0 $\mu$m spectroscopic observations of embedded high-mass YSOs in the LMC with the Infrared Camera (IRC) on board the \textit{AKARI} satellite \citep{TON07,Mur07}. 
They find  that the CO$_2$/H$_2$O ice ratio of LMC YSOs is systematically higher than those measured toward Galactic counterparts. 
This suggests that the chemical properties of circumstellar materials of YSOs may depend on the environment of the host galaxy. 
The authors suggest that intense radiation field and/or warm dust temperature in the LMC could be responsible for the different molecular abundance of ices in the LMC. 
Observations of the 15.2 $\mu$m CO$_2$ ice band toward LMC high-mass YSOs with \textit{Spitzer}/IRS suggest that a majority of CO$_2$ is locked in a water-rich (polar) ice mixture, but a fraction of CO$_2$ in a polar ice component is lower in the LMC than in our Galaxy \citep{Oli09,Sea10}. 
In addition, these studies argued that a larger number of LMC high-mass YSOs show a CO$_2$ ice profile characteristic to a pure or annealed CO$_2$ ice component compared to Galactic similar sources, suggesting a higher degree of thermal processing of ices in the LMC. 
The above previous studies suggest different physical and chemical properties of ices in the low metallicity environment, however, current studies of ices around LMC YSOs are mostly dedicated to major ice species such as H$_2$O and CO$_2$. 
Detailed information of chemically important minor ice species is therefore needed to provide a comprehensive view of ice chemistry in the low metallicity environment of the LMC.

The 3.2--3.7 $\mu$m spectral region is one of the important wavelengths for studies of solids since various C--H stretching vibrations of carbon bearing species are observed in this region. 
Solid methanol (CH$_3$OH), which has an absorption band at 3.53 $\mu$m, is an important ice mantle component and sometimes the second most abundant after water ice. 
The CH$_3$OH ice has been detected toward several YSOs in our Galaxy with an abundance ranging from 5$\%$ to 30 $\%$ relative to the water ice \citep[e.g.,][]{Gri91,Dar99,Whi11}. 
The abundance of CH$_3$OH in the solid phase is of great importance for molecular chemistry since it is believed to be a starting point for the formation of complex organic molecules in circumstellar environments of YSOs \citep[e.g., ][]{NM04,Her09}. 
However, abundances of solid CH$_3$OH in extragalactic objects remain to be investigated. 

The 3.47 $\mu$m absorption band is another interesting band in the C--H stretching region of embedded sources. 
The band is widely detected toward a variety of embedded sources such as high- to intermediate-mass YSOs \citep[e.g., ][]{All92,Bro96,Bro99,Dar01,Dar02,Ish02}, low-mass YSOs \citep[e.g., ][]{Pon03,Thi06}, and quiescent dense molecular clouds \citep[e.g., ][]{Chi96,Chi11}. 
The 3.47 $\mu$m band strength is known to correlate with the H$_2$O ice absorption depth. 
The band carrier is proposed to be interstellar nano-diamonds \citep{All92,Pir07,Men08} or ammonia hydrates formed in H$_2$O:NH$_3$ mixture ice \citep{Dar01,Dar02,Boo15}. 
However, the exact identification of the 3.47 $\mu$m band carrier is still under debate. 
Furthermore, properties of the 3.47 $\mu$m band in low metallicity environments are poorly understood.

In this paper, we present the results of L-band spectroscopic observations toward embedded high-mass YSOs in the LMC with the Infrared Spectrometer And Array Camera (ISAAC) at the Very Large Telescope (VLT) of the European Southern Observatory (ESO). 
Details of target selection, observations and data reduction are summarized in $\S$2. 
Results of observations and spectral analysis are described in $\S$3. 
Properties of solid methanol and the 3.47 $\mu$m absorption band in the LMC are discussed in $\S$4. 
Implications for the formation of organic molecules in low metallicity galaxies are also discussed in this section. 
Conclusions of this work are summarized in $\S$5.

%%%%%%%%%%%%%%%%%%%%%%%%%%%\begin{landscape}
\begin{table*}
\caption{Summary of the observations}
\label{target}
\centering
\begin{tabular}{ l c c c c c c}
\hline\hline
Source & Other ID &RA              & Dec           & [3.55]\tablefootmark{a} & t$_{int}$\tablefootmark{b}  & Standard                          \\
           &              &(J2000)        & (J2000)      & (mag)                        & (min)                            & Star\tablefootmark{c}         \\
\hline
ST1    & MSX LMC 940           & 05:39:31.15  &  -70:12:16.8  &  10.52  & 70   & H29   \\
ST2    & MSX LMC 501           & 05:22:12.56  &  -67:58:32.2  &  9.75   & 70    & H29    \\
ST3    & J052546.51-661411.5 & 05:25:46.69  &  -66:14:11.3  &  9.98   & 60    & H16    \\
ST4    & J051449.43-671221.4 & 05:14:49.41        &  -67:12:21.5  &  10.45  & 80   & H16    \\
ST5    & J053054.27-683428.2 & 05:30:54.27        &  -68:34:28.2  &  10.03  & 30   & H29    \\
ST6    & J053941.12-692916.8 & 05:39:41.08        &  -69:29:16.8  &  12.12  & 220 & H16, H29    \\
ST7    & J052351.13-680712.2 & 05:23:51.15        &  -68:07:12.2  &  10.28  & 55   & H29    \\
ST10   & MSX LMC 1229        & 04:56:40.80       &  -66:32:30.4  &  10.67  & 195 & H16, H29    \\
ST16   & MSX LMC 318          & 05:19:12.30       &  -69:09:06.8  &  9.89   & 30   & H16    \\
\hline
\textit{Archival sources} & & & & & \\
ST14   & MSX LMC 1275 & 04:58:54.33  &  -66:07:18.8  &  10.67   & 65   & H37    \\
ST17   & MSX LMC 94     & 05:10:24.15  &  -70:14:06.7  &  11.51   & 65   & H37    \\
\hline
\end{tabular}
\tablefoot{
\tablefoottext{a}{\textit{Spitzer}/IRAC band 1 (3.55 $\mu$m) magnitude after color correction.}
\tablefoottext{b}{Total on-source integration time.}
\tablefoottext{c}{H16, H29 and H37 refer to HIP16368, HIP29134 and HIP37623, respectively.}
}
\end{table*}
%%%%%%%%%%%%%%%%%%%%%%%%%%

%%%%%%%%%%%%%%%%%%%%%%%%%%%\begin{landscape}
\begin{table*}
\caption{Spectroscopic standard stars used for calibration in this study}
\label{std}
\centering
\begin{tabular}{ l c c c c c }
\hline\hline
Name & RA                       & Dec               & Spectral                       & V\tablefootmark{a}        & L'\tablefootmark{b}         \\
         & (J2000)                 & (J2000)          & Type\tablefootmark{a}   & (mag)                           & (mag)   \\
\hline
HIP16368 & 03:30:51.71     & -66:29:23.0    & B8V                            & 5.04                             & 5.20      \\
HIP29134 & 06:08:44.26     & -68:50:36.3    & B8V                            & 5.81                             & 5.80      \\
HIP37623 & 07:43:11.98     & -36:03:00.3    & B5V                            & 5.59                             & 5.99\tablefootmark{b}      \\
\hline
\end{tabular}
\tablefoot{
\tablefoottext{a}{Data taken from the SIMBAD database. }
\tablefoottext{b}{Estimated assuming [K] -- [L'] = -0.05 (ref. ISAAC web page for spectroscopic standards). }
}
\end{table*}
%%%%%%%%%%%%%%%%%%%%%%%%%%

\section{Observations and data reduction}
\subsection{Selection of targets}
Nine high-mass YSOs are selected based on our previous near-infrared spectroscopic observations with \textit{AKARI} presented in \citet{ST10} and \citet{thesis}. 
In addition, we obtained ISAAC spectroscopic data from the ESO archive for two high-mass YSOs (089.C-0882(C); PI: J. M. Oliveira) whose \textit{AKARI} spectra are also presented in \citet{thesis}. 
All of these YSOs show  H$_2$O and CO$_2$ ice absorption bands in their near-infrared spectra, suggesting that they are appropriate targets to investigate minor ice species. 
Tentative detections of solid CH$_3$OH, CO, and XCN bands were reported for several \textit{AKARI} samples in our previous study, but these were less conclusive due to the low spectral resolution (R $\sim$ 80) and low spatial resolution. 
We selected sources with relatively large ice column densities and high L-band fluxes for the present observations. 
Details of the targets and observations are summarized in Table \ref{target}.

\subsection{VLT/ISAAC L-band spectroscopy}
Spectra of nine high-mass YSOs were obtained between January 21 and 28, 2013, at the VLT UT3 using the infrared spectrograph ISAAC. 
Observations were carried out as a part of the ESO normal program 090.C-0497(A) (PI: E. Dartois). 
The VLT, located on Cerro Paranal, Chile, has a primary mirror of 8.2 m in diameter. 
The ISAAC was installed at the Nasmyth A focus of VLT/UT3 at the time of our observations. 

We conducted L-band slit spectroscopy with a low-resolution grating, which covers 2.8 $\mu$m to 4.2 $\mu$m with the 120$\arcsec$ length slit. 
The central wavelength of the grating was set to 3.55 $\mu$m. 
The pixel scale of the detector is 0.1484$\arcsec$ and the size is 1024 $\times$ 1024. 
Target acquisition was carried out with the ESO L-band filter centered at 3.78 $\mu$m. 
The atmospheric seeing was typically 0.6$\arcsec$--0.7$\arcsec$ during our observation nights. 
The slit width was set to 0.6$\arcsec$ for every target except ST5, whose slit width was set to 0.3$\arcsec$ considering its brightness and atmospheric conditions of the night. 
The slit width for archival sources (ST14 and ST17) is 1$\arcsec$. 
The achieved spatial resolution in actual physical scale at the distance of the LMC is 0.15 pc and 0.07 pc for the 0.6$\arcsec$ and 0.3$\arcsec$ slit, respectively. 
The resultant spectral resolution is R $\sim$600 for the 0.6$\arcsec$ slit on raw data. 
The actual spectral resolution of fully calibrated data is slightly worse due to spectral binning (see $\S$2.3). 
Sky background emission was canceled out by chopping and nodding with ABBA sequences, in which each target was observed at two different positions along the slit, and the sky was removed by subtracting one frame from the other. 
The chopping throw and direction were carefully determined by inspecting the \textit{AKARI}/IRC 3.2 $\mu$m images \citep{Kat12} or the \textit{Spitzer}/IRAC 3.55 $\mu$m images \citep{Mei06} at each target position. 
The adopted chopping throw ranges from 5$\arcsec$ to 15$\arcsec$ depending on the targets. 
Standard stars were observed for calibration purpose before and after observations of each target. 
Properties of the standard stars are summarized in Table \ref{std}. 
Airmass differences between targets and standard stars were typically smaller than 0.1.

%%%%%%%%%%
\begin{figure*}[!]
\begin{center}
\includegraphics[width=15.5cm, angle=0]{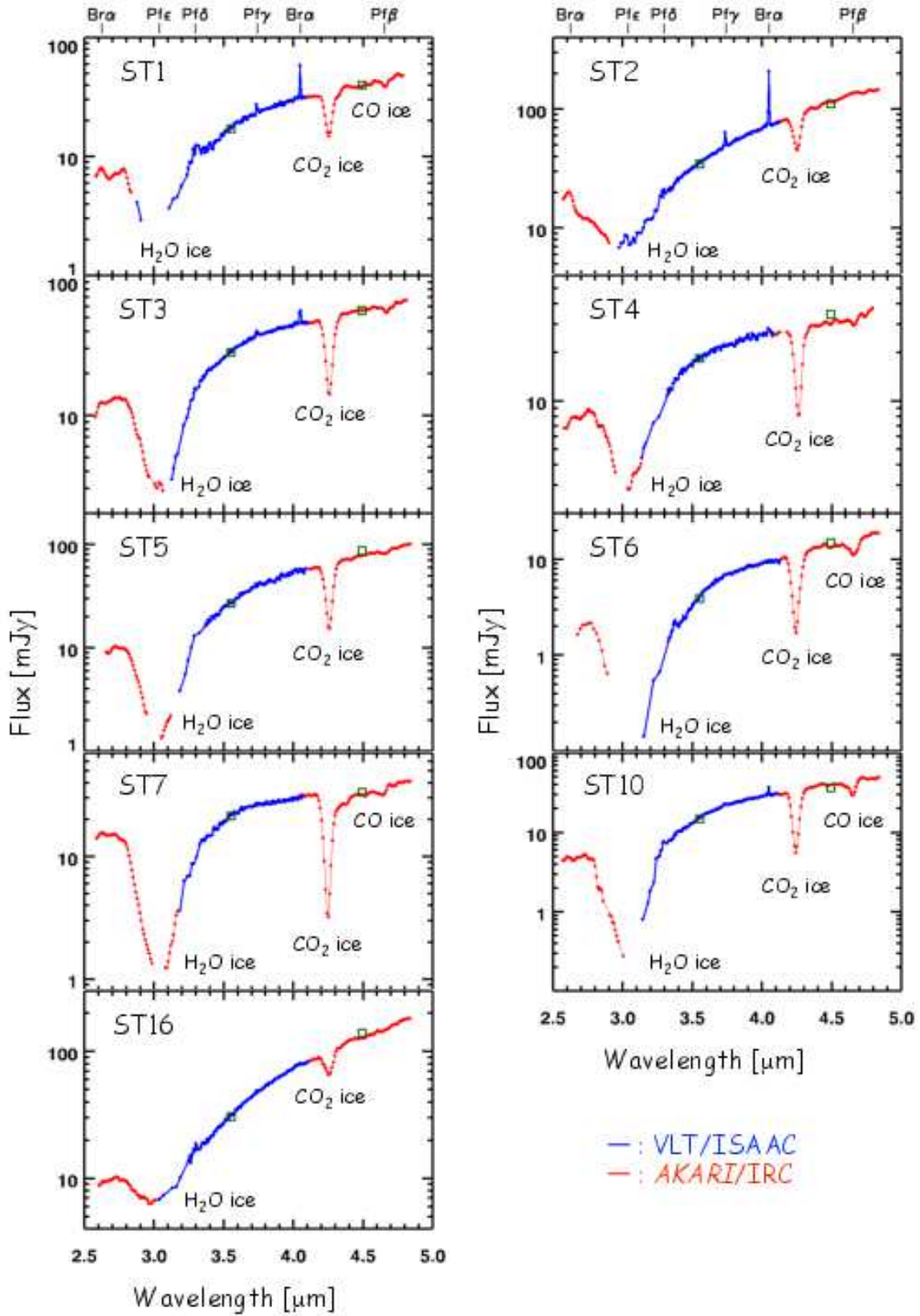}
\caption{VLT/ISAAC and {\it AKARI}/IRC 2.5--5 $\mu$m spectra of embedded high-mass YSOs in the LMC. 
The blue circles connected by thick lines represent the VLT L-band spectra obtained in this work. 
The red circles connected by thin lines represent the {\it AKARI} spectra (the original data taken from \citet{ST10} and \citet{thesis}). 
The green open squares represent photometric fluxes of {\it Spitzer}/IRAC band 1 (3.55 $\mu$m) and band 2 (4.5 $\mu$m) measured in the SAGE survey. 
Detected ice absorption bands are labeled in each panel and the wavelengths of hydrogen recombination lines are indicated in the upper panels. 
A color version of this figure is available in the online journal.
}
\label{Spec}
\end{center}
\end{figure*}

\begin{figure*}[!]
\begin{center}
\includegraphics[width=15.5cm, angle=0]{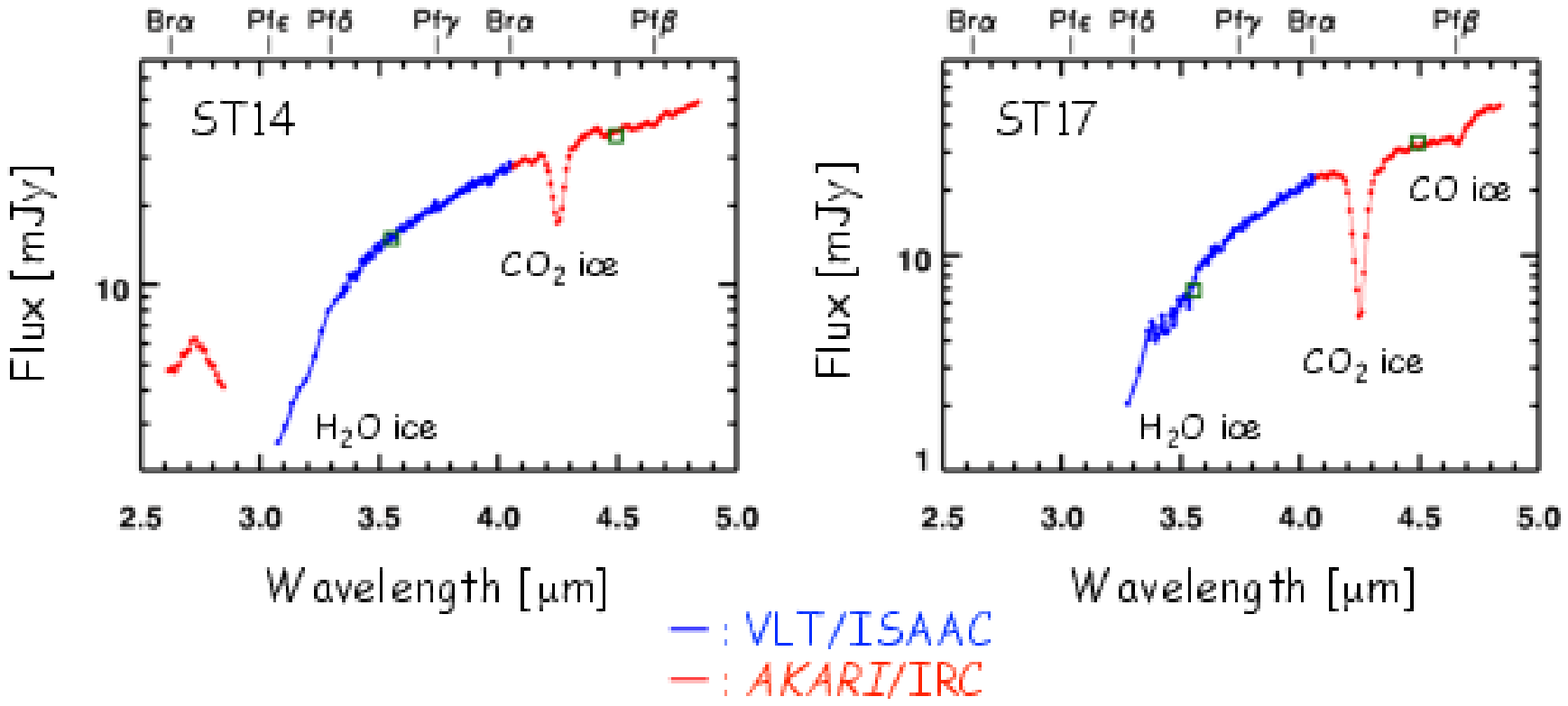}
\caption{Same as Fig. \ref{Spec}, for archival sources. 
}
\label{Spec_add}
\end{center}
\end{figure*}
%%%%%%%%%%

\subsection{Data reduction}
The raw data are reduced using our own IDL-based programs developed for VLT/ISAAC slit spectroscopy data. 
Data of two archival sources are reduced in the same manner as our target sources. 
First, individual exposure data are coadded to cancel bad pixels and to increase the signal-to-noise ratio. 
Since the present spectroscopic observations were performed with the ABBA chopping sequences, spectra of objects are detected at three positions along the chopping direction on coadded data (one positive spectrum at the ON position and two negative spectra at the OFF positions). 
To extract each spectrum, we measure the central position and FWHM of signals along the spatial direction at multiple wavelength points. 
We then extract the three spectra using the extraction width and position determined above and combine them into one spectrum. 
Uncertainties of data are estimated based on the signals at sky positions, which are typically located  at 15 pixels apart from the source position. 
We extract spectra of targets and standard stars in the same manner. 
The extracted spectra are smoothed by a Savitzky-Golay filter with an appropriate polynomial degree and filter width. 

To obtain flux calibrated spectra, the target spectra are divided by the spectrum of the associated standard star. 
The wavelength calibration is carried out  using telluric absorption lines before this division, and slight wavelength shifts between target's and standard star's spectra are corrected. 
We also correct airmass differences between targets and standard stars. 
A theoretical spectrum of the standard stars is applied to the spectra for flux calibration. 
We here assume blackbody spectra for the standard stars. 
The temperature of the blackbody is estimated by fitting the Planck function to photometric fluxes in BVIJHK bands, which are taken from the SIMBAD database \footnote{http://simbad.u-strasbg.fr/simbad/}. 
In addition, we add photospheric absorption lines of hydrogen (Pf $\zeta$ at 2.873 $\mu$m, Pf $\epsilon$ at 3.039 $\mu$m, Pf $\delta$ at 3.297 $\mu$m, Pf $\gamma$ at 3.741 $\mu$m, Br $\alpha$ at 4.0523 $\mu$m), which intrinsically exist in the spectrum of standard stars. 
The strength and width of the absorption lines are measured by fitting a Gaussian to the Br $\alpha$ line, which is usually the most prominent and less contaminated with telluric lines. 
Then, the strengths of other absorption lines relative to Br $\alpha$ are determined based on a Kurucz stellar atmosphere model (T$_{eff}$ = 11900 K, log $g$ = +4.04, [Fe/H] = 0), which is taken from The 1993 Kurucz Stellar Atmospheres Atlas. 

The reduced spectra are fine-tuned with further wavelength calibration, absolute flux calibration, and appropriate smoothing as necessary. 
For sources that show emission lines of Pf $\gamma$ and Br $\alpha$, the wavelength solutions are slightly modified by interpolating the peak positions of these two emission lines. 
A geocentric radial velocity of 265 km/s is assumed to estimate the peak wavelength of hydrogen recombination lines on the sky \citep[the velocity is estimated based on Fig. 7 in][]{Fuk08}. 
The adopted velocity may have uncertainty by several tens of km/s depending on the location of sources in the LMC, but such a small velocity difference does not significantly change the wavelength solution because the present spectra are medium resolution. 
The overall wavelength accuracy of the calibrated spectra is estimated to be 0.004 $\mu$m. 
The absolute fluxes are calibrated by scaling each spectrum to the corresponding \textit{Spitzer}/IRAC band 1 photometric fluxes at 3.55 $\mu$m, which are taken from the SAGE catalog \citep{Mei06}. 
The color-correction is applied to the photometric data to accurately determine the in-band fluxes. 
The appropriate color correction factors are estimated based on the gradient of each spectrum between 3.2 and 3.9 $\mu$m and the tables of color correction factors given in the IRAC instrument handbook. 

Finally, the spectra are binned to achieve the S/N ratio that is necessary for the subsequent spectral analysis. 
Noisy spectral regions due to telluric lines and bad pixels are carefully inspected and masked before data binning. 
The resultant spectral resolution of the present VLT/ISAAC data is $\lambda$/$\Delta$$\lambda$ $\sim$500 for the 3.2--4.0 $\mu$m region. 
The spectral resolution shorter than 3.2 $\mu$m is worse by a factor of two to four since more data points are binned in this wavelength region because of increased noise. 

Calibrated VLT/ISAAC spectra are shown in Figs. \ref{Spec}--\ref{Spec_add}. 
Since broad wavelength coverage is crucial for reliable continuum determination in the subsequent spectral analysis, \textit{AKARI}/IRC grism spectra ($\lambda$/$\Delta$$\lambda$ $\sim$100, data taken from \citet{ST10} and \citet{thesis}) are concatenated to the short and long wavelength sides of the ISAAC spectra. 
The IRC spectra are scaled to match the ISAAC spectra around 3 $\mu$m and 4 $\mu$m (see also Fig. \ref{App_Spec} for comparison of ISAAC and IRC spectra). 
The concatenated \textit{AKARI} spectra are also shown in the figures.

%%%%%%%%%%
\begin{figure*}[!]
\begin{center}
\includegraphics[width=18cm, angle=0]{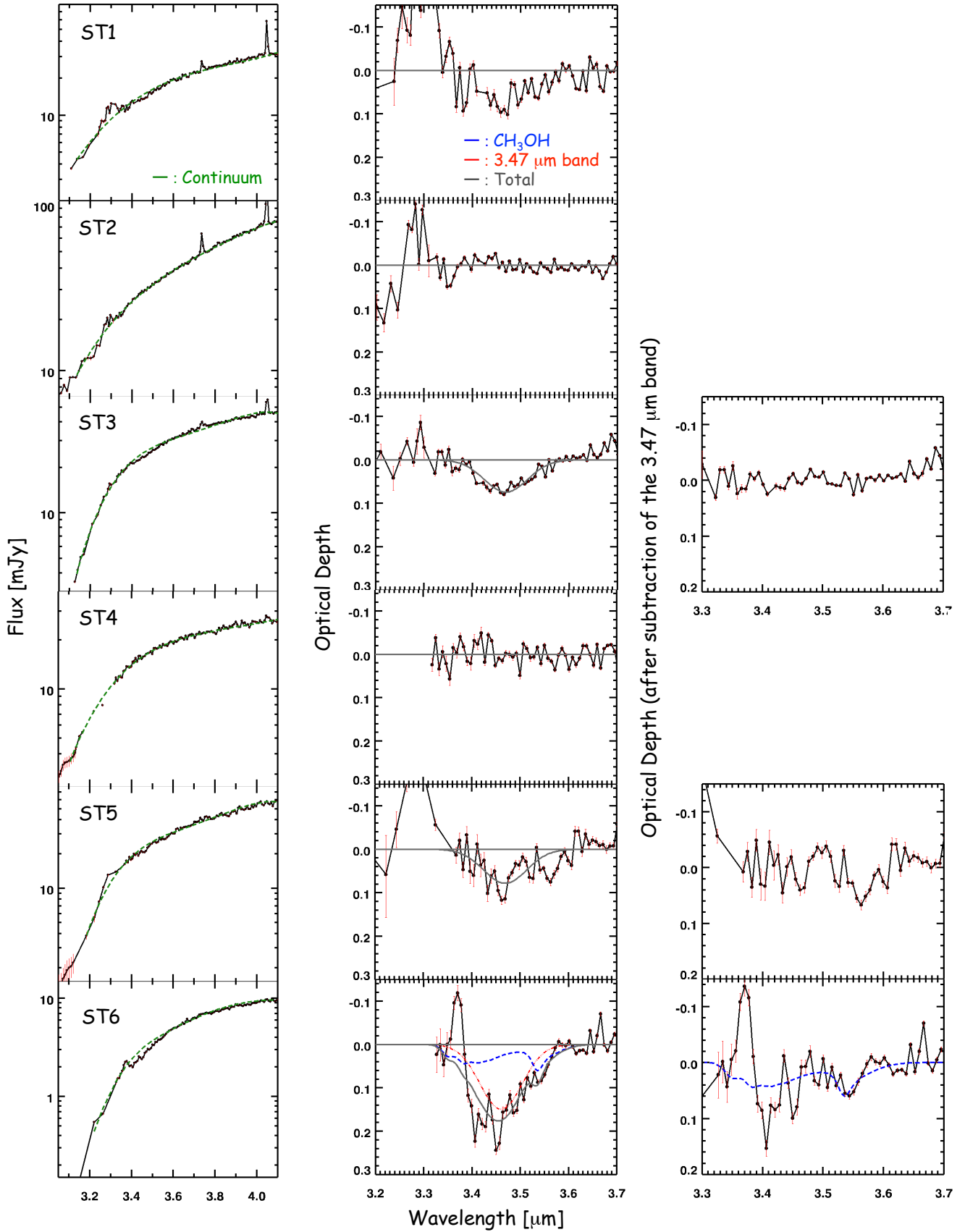}
\caption{
Results of the spectral analysis for the 3.47 $\mu$m and 3.53 $\mu$m (CH$_3$OH) absorption bands. 
Left: derived continua (dashed lines, green) are shown with the observed spectra. 
Middle: results of the spectral fitting. 
The dashed lines (blue) represent the laboratory CH$_3$OH ice spectrum and the dot-dashed lines (red) represent the 3.47 $\mu$m absorption band. 
The solid lines (gray) show the sum of those two components. 
Right: residual spectra after subtraction of the 3.47 $\mu$m band. The spectra are only shown  for the sources in which the 3.47 $\mu$m band is detected. 
}
\label{Fit1}
\end{center}
\end{figure*}

\begin{figure*}[!]
\begin{center}
\includegraphics[width=18cm, angle=0]{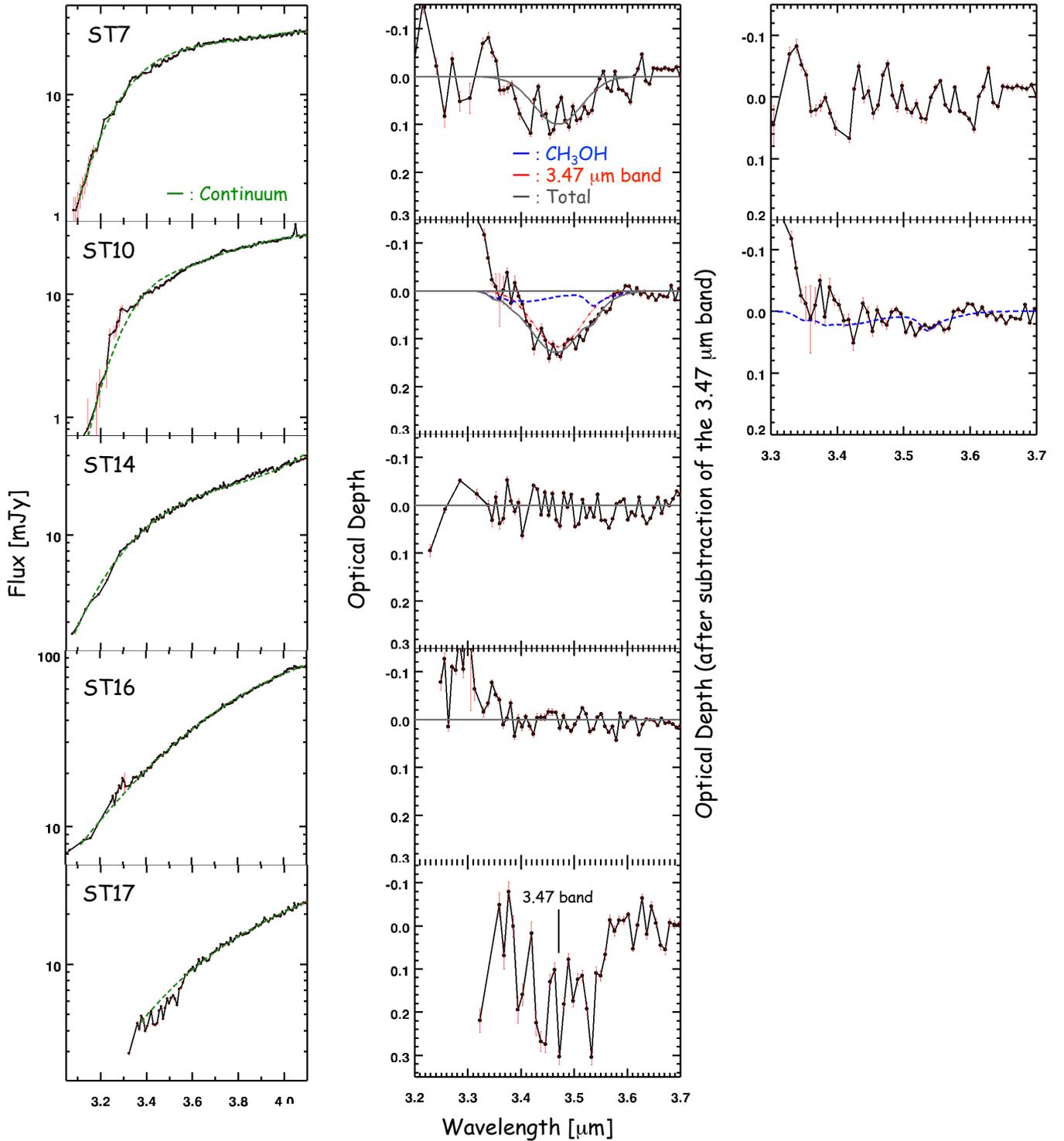}
\caption{{\it Continued}
}
\label{Fit2}
\end{center}
\end{figure*}
%%%%%%%%%%

%%%%%%%%%%
\begin{figure*}[!]
\begin{center}
\includegraphics[width=15.5cm, angle=0]{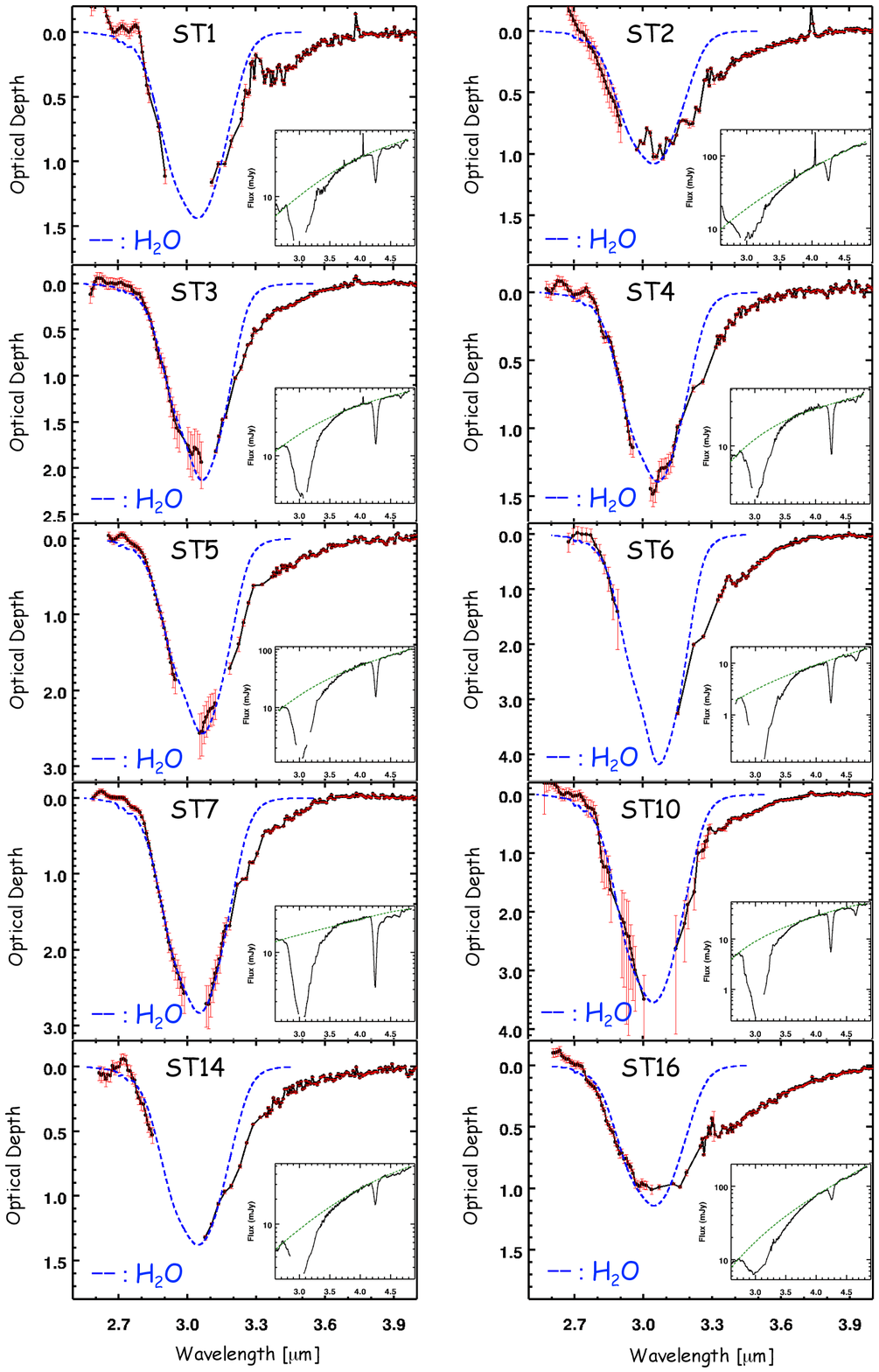}
\caption{Results of the spectral analysis for the 3.05 $\mu$m H$_2$O ice band. 
The dashed lines (blue) represent the laboratory H$_2$O ice spectrum fitted to the observed spectrum (solid lines, black). 
Derived continua are shown in the lower right side of each panel with  dashed lines (green). 
}
\label{WFit1}
\end{center}
\end{figure*}
%%%%%%%%%%

\section{Results} 
\subsection{Observed spectra} 
Figures \ref{Spec}--\ref{Spec_add} show the VLT/ISAAC spectra of high-mass embedded YSOs in the LMC together with their \textit{AKARI}/IRC spectra. 
The 3.05 $\mu$m H$_2$O ice absorption band (O--H stretching mode) is detected toward all of the eleven sources.  
Since the short-wavelength side of the water absorption is not detected owing to low atmospheric transmission, the present L-band spectroscopy mainly detects the long-wavelength side of the water ice absorption; i.e., the red wing of  the water ice absorption \citep[e.g.,][]{Dar01,Dar02}. 
The 3.53 $\mu$m CH$_3$OH ice absorption band (C--H stretching mode) is detected toward ST6 and ST10, and the strength of the absorption is found to be weak. 
The band is absent toward all the other sources, but meaningful upper limits are obtained. 
The 3.47 $\mu$m absorption band, which is often detected toward embedded sources in our Galaxy, is detected toward six LMC sources (ST3, ST5, ST6, ST7, ST10, ST17). 
For ST17, however, the S/N ratio is poor. 
The absorption band around 3.4 $\mu$m, which is due to aliphatic hydrocarbon and detected toward diffuse interstellar clouds, is not present in our observations. 
Details of the absorption bands in the 3.4--3.5 $\mu$m region are discussed in $\S$4 after careful baseline correction of continuum. 
Hydrogen recombination lines (Pf $\gamma$ at 3.7406 $\mu$m and Br $\alpha$ at 4.0523 $\mu$m) are also detected in several sources. 

The VLT and \textit{AKARI} spectra are compared in Fig. \ref{App_Spec}. 
Each panel in the figure shows three spectra including, a VLT spectrum before spectral smoothing, a VLT spectrum after spectral smoothing and concatenation with \textit{AKARI} data, and an original \textit{AKARI} spectrum after flux level adjustment. 
Overall shapes are consistent between VLT and \textit{AKARI} data except for the region of PAH emission bands around 3.3 $\mu$m. 
The PAH emission is significantly weaker in our VLT spectra. 
Considering a different spatial resolution of VLT ($\sim$0.6$\arcsec$ = 0.15 pc at the LMC) and \textit{AKARI} ($\sim$6$\arcsec$ = 1.5 pc), this suggests that the PAH emission mainly arises from an extended region ($>$ 1 pc) surrounding our target YSOs, while continuum emission and ice absorption arise from a compact region ($\sim$0.1 pc).

%%%%%%%%%%%%%%%%%%%%%%%%%%%\begin{landscape}
\begin{table}[b]
\centering
\caption{Optical depths of the observed absorption bands for high-mass YSOs in the LMC}
\label{Tab_tau}
\begin{tabular}{ l c c c }
\hline\hline
Source  &  $\tau$$_{3.05 \mu m}$    &  $\tau$$_{3.47 \mu m}$    &  $\tau$$_{3.53 \mu m}$  \\
\hline
ST1      &  1.44 $\pm$ 0.19             &  $<$0.04                        &  $<$0.02                     \\
ST2      &  1.08 $\pm$ 0.09             &  $<$0.02                        &  $<$0.02                      \\
ST3      &  2.19 $\pm$ 0.04             &  0.07 $\pm$ 0.01            &  $<$0.01                       \\
ST4      &  1.39 $\pm$ 0.11             &  $<$0.02                        &  $<$0.02                       \\
ST5      &  2.57 $\pm$ 0.29             &  0.08 $\pm$ 0.03            &  $<$0.04                        \\
ST6      &  4.18 $\pm$ 0.58             &  0.15 $\pm$ 0.02            &  0.06 $\pm$ 0.02             \\
ST7      &  2.83 $\pm$ 0.12             &  0.10 $\pm$ 0.02            &  $<$0.04                         \\
ST10    &  3.54 $\pm$ 0.60             &  0.11 $\pm$ 0.01            &  0.04 $\pm$ 0.01             \\
ST14    &  1.38 $\pm$ 0.07             &  $<$0.03                        &  $<$0.03                          \\
ST16    &  1.14 $\pm$ 0.19             &  $<$0.02                        &  $<$0.02                          \\
ST17    &  ...                                    &  0.2 $\pm$ 0.1              &  ...                                  \\
\hline
\end{tabular}
\tablefoot{Uncertainties and upper limits are of 3 $\sigma$ level and do not include systematic errors due to continuum determination. }
\end{table}
%%%%%%%%%%%%%%%%%%%%%%%%%%

\subsection{Spectral analysis of absorption bands} 
The observed spectra are used to derive column densities of detected molecular species. 
We derive column densities of the H$_2$O and CH$_3$OH ices by fitting laboratory ice spectra to the observed spectra after careful subtraction of continuum. 
The optical depth of the 3.47 $\mu$m absorption band is also estimated in conjunction with spectral fitting of the CH$_3$OH ice band. 
For ST17, we analyze only the 3.2--3.7 $\mu$m region because of lack of spectral data shorter than 3.2 $\mu$m. 
Details of continuum subtraction and spectral fitting for the 3.2--3.7 $\mu$m region and the water ice band are described below.

\subsubsection{Continuum determination: 3.2--3.7 $\mu$m region} 
Continuum baselines are carefully estimated to accurately derive the optical depth of the 3.47 $\mu$m and 3.53 $\mu$m absorption bands, since it is located in the red wing of the deep water ice absorption. 
In our analysis, local continuum is derived by fitting a three- to fourth-order polynomial function to the selected continuum regions. 
The wavelength regions used for the local continuum are typically 3.09--3.15 $\mu$m, 3.35--3.39 $\mu$m, 3.59--3.72 $\mu$m, 3.78--3.81 $\mu$m, 3.98--4.01 $\mu$m, and 4.07--4.15 $\mu$m. 
The spectral region of PAH emission is not used for the continuum determination. 
Each adopted continuum is shown in the left panel of Figs. \ref{Fit1}--\ref{Fit2}. 
These local continua mimic the smooth profile of the water ice wing and enable us to extract the buried absorption component. 
Observed spectra are divided by these continua and converted to optical depth (right panel of Fig. \ref{Fit1}--\ref{Fit2}). 
A similar definition of continuum is adopted in \citet{Bro99} and \citet{Dar99}, which investigate the 3.47 $\mu$m and 3.53 $\mu$m absorption band for Galactic high-mass YSOs. 
Since continuum subtraction significantly affects the absorption depth of the 3.53 $\mu$m band, we employ the similar continuum definition and compare absorption depths of LMC sources with those of the Galactic sources that are reported in the above-mentioned papers.

\subsubsection{Spectral fitting: 3.2--3.7 $\mu$m region} 
We fit a laboratory ice spectrum to the observed spectra to derive ice column densities. 
Since the 3.53 $\mu$m CH$_3$OH band partially overlaps with the 3.47 $\mu$m absorption band, these two absorption components are simultaneously fitted with a $\chi^2$ minimization method. 
Wavelength regions between 3.3 $\mu$m and 3.7 $\mu$m are used for the fit. 
For the CH$_3$OH band, we fit a laboratory profile of the pure CH$_3$OH ice at 10 K, which is taken from the Leiden Molecular Astrophysics database\footnote{http://www.strw.leidenuniv.nl/lab/databases/} \citep{Ger95,Ger96}. 
We tried to fit various CH$_3$OH ice mixtures in this analysis and we confirm through visual inspection that the pure CH$_3$OH ice results in the best fit among the investigated ice mixtures. 
It is reported that the profile of the pure CH$_3$OH ice also fits well to the 3.53 $\mu$m band of Galactic high-mass YSOs \citep{All92,Bro99}. 
For the 3.47 $\mu$m band, we fit a Gaussian profile ($\lambda$$_c$ = 3.469 $\mu$m, FWHM = 0.105 $\mu$m) except for ST6. 
The assumed profile is consistent with the typical profile of the 3.47 $\mu$m band observed toward Galactic high-mass YSOs \citep{Bro99}. 
For ST6, we use another Gaussian profile ($\lambda$$_c$ = 3.460 $\mu$m, FWHM = 0.103 $\mu$m) for the fitting since the former profile results in a poor fit. 
This kind of blueshifted profile is relatively rare but observed toward a Galactic high-mass YSO Mon R2 IRS 2 \citep{Bro99}. 

Based on the above fitting, we derive the peak optical depth of the 3.53 $\mu$m CH$_3$OH band and then calculate ice column densities using the following equation: 
\begin{equation}
N(CH_3OH) = \tau(3.53)  \Delta\nu / A, \label{Eq_column}
\end{equation}
where $N(CH$$_3$$OH)$ is the column density of the CH$_3$OH ice in units of cm$^{-2}$,  $\tau(3.53)$ is the peak optical depth of the 3.53 $\mu$m absorption band as derived by our spectral fitting, $\Delta$$\nu$ is the FWHM of the absorption band, and $A$ is the band strength as measured in the laboratory. 
We adopt the band strengths of the pure CH$_3$OH ice as 5.3$\times$10$^{-18}$ cm molecule$^{-1}$ \citep{Ker99}, and $\Delta$$\nu$ is assumed to be 31 cm$^{-1}$ \citep{Sch96}. 
The result of the fitting is shown in the right panel of Figs. \ref{Fit1}--\ref{Fit2}. 
Derived optical depths and column densities are summarized in Tables \ref{Tab_tau} and \ref{Tab_column}.

%%%%%%%%%%%%%%%%%%%%%%%%%%%\begin{landscape}
\begin{table*}
\centering
\caption{Derived ice column densities and abundances for high-mass YSOs in the LMC}
\label{Tab_column}
\begin{tabular}{ l c c c c c c c}
\hline\hline
Source  &  \textit{N}(H$_2$O)     & \textit{N}(CH$_3$OH) & \textit{N}(CO)                 &  \textit{N}(CO$_2$)       & \textit{N}(CH$_3$OH)          &  \textit{N}(CO) &  \textit{N}(CO$_2$)                     \\
 &  &  &  &  & /\textit{N}(H$_2$O) & /\textit{N}(H$_2$O) & /\textit{N}(H$_2$O) \\
             & (10$^{17}$ cm$^{-2}$)& (10$^{17}$ cm$^{-2}$) &  (10$^{17}$ cm$^{-2}$) &  (10$^{17}$ cm$^{-2}$) &   ($\%$)  &  ($\%$ ) &  ($\%$ )  \\
\hline
ST1      &  24.88 $\pm$ 3.29        &  $<$1.2                       &  4.1 $\pm$ 1.1   &  5.9 $\pm$ 0.5                              &  $<$4.8    &   16.5 &   23.7  \\
ST2      &  18.65 $\pm$ 1.58        &  $<$1.2                       &  $<$0.5              &  3.8 $\pm$ 0.4                              &  $<$6.4    &   <3    &   20.4 \\
ST3      &  34.43 $\pm$ 0.66        &  $<$0.6                       &  1.5 $\pm$ 0.2\tablefootmark{a} &  10.1 $\pm$ 1.3   &  $<$1.7    &   4.5   &   29.3  \\
ST4      &  23.22 $\pm$ 1.80        &  $<$1.2                       &  1.6 $\pm$ 0.1\tablefootmark{a} &  8.4 $\pm$ 1.1     &  $<$5.2    &   6.9   &   36.2  \\
ST5      &  40.60 $\pm$ 4.64        &  $<$2.3                       &  1.9 $\pm$ 0.8   &  11.4 $\pm$ 0.9                              &  $<$5.7   &   4.7   &   28.1 \\
ST6      &  62.10 $\pm$ 8.62        &  3.5 $\pm$ 1.2            &   $<$15              &  21.5 $\pm$ 6.2                              &  5.6        &   <24  &   34.6 \\
ST7      &  48.02 $\pm$ 2.10        &  $<$2.3                       &  2.6 $\pm$ 0.2\tablefootmark{a} &  26.0 $\pm$ 2.9     &  $<$4.8  &   5.4   &   54.1 \\
ST10    &  61.42 $\pm$ 10.36      &  2.3 $\pm$ 0.6            &  9.8 $\pm$ 3.3   &  17.9 $\pm$ 3.4                              &  3.7         &   16.0 &   29.1 \\
ST14    &  23.82 $\pm$ 1.24        &  $<$1.8                       &  2.8 $\pm$ 0.9   &  3.8 $\pm$ 0.3                                &  $<$7.6   &  11.8  &   16.0 \\
ST16    &  19.64 $\pm$ 3.21        &  $<$1.2                       &  $<$2                 &  2.7 $\pm$ 0.2                                &  $<$6.1   &  <10   &   13.8  \\
ST17    &  ...                                 &   ...                              &  5.5 $\pm$ 1.7   &  14.3 $\pm$ 1.8                              &  ...           &  ...      &   ...       \\
\hline
\end{tabular}
\tablefoot{Uncertainties and upper limits are of 3 $\sigma$ level and do not include systematic errors due to continuum determination and adopted band strengths.
Ref. \tablefoottext{a}{\citet{Oli11}}
}
\end{table*}
%%%%%%%%%%%%%%%%%%%%%%%%%%

\subsubsection{Continuum determination: H$_2$O, CO, and CO$_2$ ices} 
For H$_2$O , CO, and CO$_2$ ices, we use global continuum that is derived by fitting a polynomial of the third or fourth order to the observed spectra. 
The wavelength regions used for the global continuum are typically 2.6--2.7 $\mu$m, 4.1--4.15 $\mu$m, and 4.8--4.9 $\mu$m, which are set to avoid spectral regions that show prominent absorption or emission features. 
The estimated continua are shown in Fig. \ref{WFit1} together with the optical depth spectra derived based on these global continuum.

\subsubsection{Spectral fitting: H$_2$O, CO, and CO$_2$ ices} 
Although H$_2$O, CO, and CO$_2$ ice column densities of the present targets are derived in our previous studies \citep{ST10,thesis}, we re-evaluate their column densities using the combined VLT + \textit{AKARI} spectra obtained in this work. 
The contamination by PAH emission bands and diffuse warm dust emission is significantly reduced in the present VLT spectra, thanks to the higher spatial resolution, which enables us to derive more reliable ice column densities. 

For water ice, we fit laboratory ice spectra to the observed spectra with a $\chi^2$ minimization method and derived the column density by the following equation: 
\begin{equation}
N(H_2O) = \int \tau  d\nu / A, \label{Eq_column}
\end{equation}
where $N(H$$_2$$O)$ is the column density of the H$_2$O ice in units of cm$^{-2}$, $\tau$ is the optical depth, $\nu$ the wavenumber in units of cm$^{-1}$, $A$ is the band strength based on laboratory measurements. 
The integration is performed over the wavelength region of the H$_2$O ice absorption band. 
Wavelength regions used for the fit are set to avoid the bottom and the long-wavelength wing of the H$_2$O absorption band. 
The fitted wavelengths are typically 2.75--2.95 $\mu$m and 3.1--3.15 $\mu$m. 
Laboratory ice profiles of the H$_2$O and CO$_2$ ice mixture at 10 K and 80 K (H$_2$O:CO$_2$ = 100:14), which are similar to the typical chemical compositions of circumstellar ices around high-mass YSOs, are simultaneously fitted to the observed spectra. 
The laboratory spectrum is taken from the Leiden database. 
We adopt the band strength of the H$_2$O ice band as 2.0$\times$10$^{-16}$ cm molecule$^{-1}$ \citep{Ger95}. 
The result of the fitting is shown in Fig. \ref{WFit1} and derived column densities are summarized in Tables \ref{Tab_tau} and \ref{Tab_column}. 
The resultant H$_2$O ice column densities are somewhat larger than those derived by \textit{AKARI} data in \citet{ST10}, particularly for ST2, ST5, ST7, and ST10. 
This is because ISAAC spectra are less contaminated by surrounding emissions by PAH and warm dust. 

For CO and CO$_2$ ices, we use the same method as presented in \citet{ST10} for spectral fitting and for calculation of ice column densities. 
The results of spectral fitting are shown in Fig. \ref{App_Spec2}. 
The derived column densities are summarized in Tables \ref{Tab_column}. 
The resultant ice column densities are generally larger than the previous estimates due to a decrease in the overall flux level of \textit{AKARI} spectra.

\subsubsection{Notes on individual sources} 
\textbf{ST6:} The source shows the deepest absorptions of the H$_2$O, CH$_3$OH, and 3.47 $\mu$m band among the sources examined in this study.  
This suggests the deeply-embedded and chemically-rich nature of the source. 
We additionally analyzed \textit{Spitzer}/IRS 5--33 $\mu$m spectrum of the source to investigate properties of ice absorption bands in the mid-infrared region. 
The IRS spectral data were taken from the SAGE-Spec database \citep{Kem10,Woo11}. 
The result of the analysis is presented in $\S$ 3.3. 

\textbf{ST10:} The source shows the second deepest absorptions of the H$_2$O, CH$_3$OH, and 3.47 $\mu$m band after ST6. 
We examined the IRS spectrum of the source, but it was severely contaminated by PAH emissions. 

\textbf{ST17:} The source shows absorptions of the H$_2$O and 3.47 $\mu$m band, although the S/N ratio of the spectrum is poor. 
The optical depth of the 3.47 $\mu$m band is estimated by visual inspection, but the uncertainty is large since the bottom of the absorption band is noisy. 
The source shows a hint of the CH$_3$OH ice band, but it is difficult to claim the detection with this S/N. 
The \textit{AKARI} spectrum of the source is contaminated by a nearby source particularly in the wavelength region shorter than 4 $\mu$m. 
Thus the short-wavelength side of the H$_2$O absorption band is not available for this source, which makes it difficult to estimate the column density of the H$_2$O ice. 
Data of ST17 are not used in the following discussion section because of the low spectral quality.

%%%%%%%%%%
\begin{figure*}[!]
\begin{center}
\includegraphics[width=18cm, angle=0]{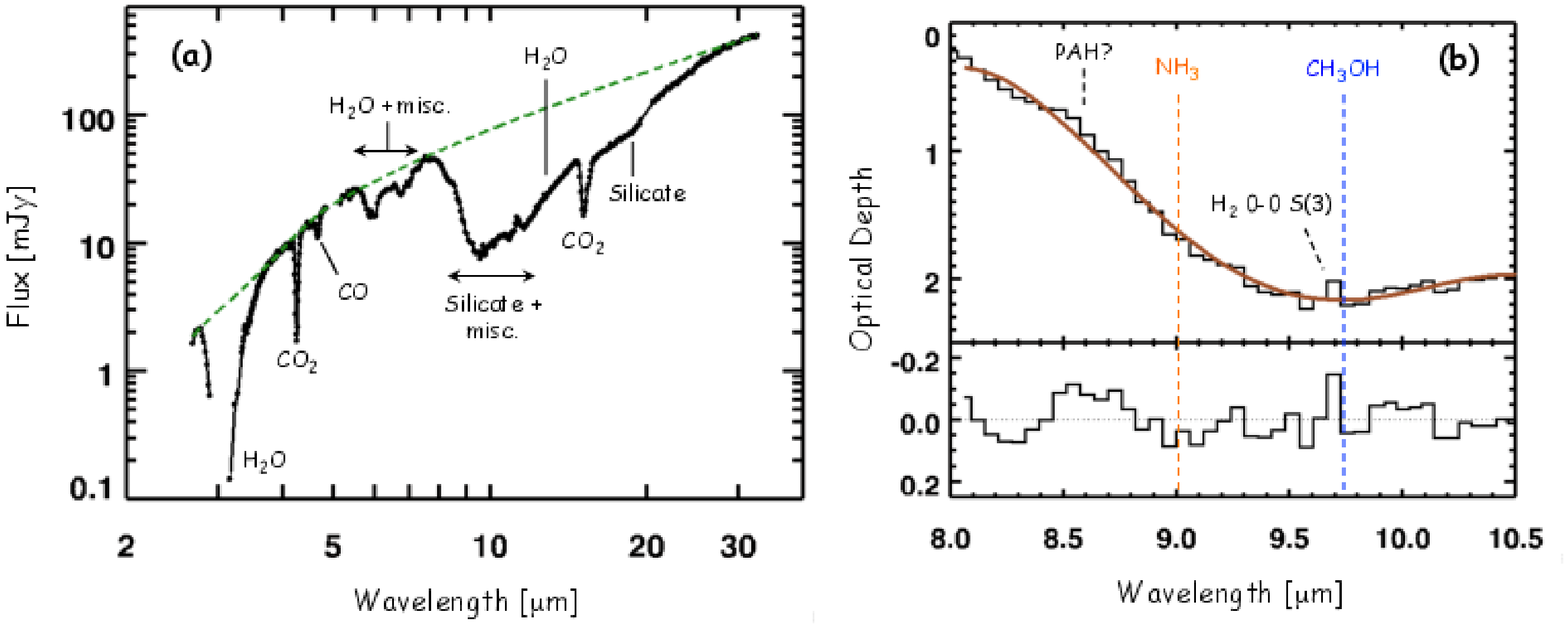}
\caption{
(a) \textit{AKARI}/IRC 2.5--5 $\mu$m and \textit{Spitzer}/IRS 5--33 $\mu$m spectrum of ST6. 
The derived continuum is shown with  dashed lines (green). 
Detected absorption bands of solids are labeled. 
(b) Upper panel: optical depth spectrum of ST6 in the 8--10.5 $\mu$m region. 
The smooth solid line (brown) indicates the fitted profile of the silicate absorption band. 
Lower panel: residual spectrum after the subtraction of the silicate absorption band. 
Absorption bands of the NH$_3$ ice at 9.0 $\mu$m and  CH$_3$OH ice at 9.75 $\mu$m are not seen in the spectrum. 
}
\label{ST6_IRCIRS}
\end{center}
\end{figure*}
%%%%%%%%%%

\subsection{Ice absorption bands in the mid-infrared spectrum of ST6} 
Mid-infrared spectral regions provide us with another piece of important information about ice mantle compositions that is complementary to the near-infrared spectral data. 
Figure \ref{ST6_IRCIRS}a shows 2.5--33 $\mu$m spectrum of ST6, which shows the richest ice absorption bands in the infrared region among the sources we investigated. 
We examined IRS data of the other sources, but those are severely contaminated by surrounding emissions by PAH and warm dust. 
Detailed analysis of minor ice absorption bands is thus difficult except for those of 15.2 $\mu$m CO$_2$ ice band, which were already presented in previous studies \citep{Oli09,Sea10}. 
The IRS spectrum of ST6 is presented in \citet{Mat14}, but we carried out the analysis of ice absorption bands. 
We estimate the continuum emission by fitting a spline function to the wavelength regions that are free from absorption/emission features (2.7 $\mu$m, 4 $\mu$m, 5--5.5 $\mu$m, 7.5 $\mu$m, and 30 $\mu$m), and derive optical depth spectrum. 

Figure \ref{ST6_IRCIRS}b shows the optical depth spectrum of ST6 in the 8--10.5 $\mu$m range. 
To estimate the contribution of the NH$_3$ ice absorption band at 9.0 $\mu$m (umbrella mode), we fit and subtract the silicate absorption, and a residual spectrum is shown in the lower panel of the figure. 
Following the method described in \citet{Bot10}, the profile of the silicate absorption band is approximated by a polynomial function. 

The NH$_3$ ice 9.0 $\mu$m absorption band is not detected prominently in the residual spectrum. 
However, it seems that the absorption band is partially contaminated by the PAH emission at 8.61 $\mu$m. 
We place an upper limit of 0.1 for the peak optical depth of the NH$_3$ ice absorption band in ST6. 
This optical depth corresponds to the NH$_3$ column density of $\sim$3$\times$10$^{-17}$ cm$^{-2}$, if we assume the band strength of A = 1.3$\times$10$^{-17}$ cm molecule$^{-1}$ and the FWHM of $\Delta$$\lambda$ = 0.3 $\mu$m \citep{Bot10}. 
Thus, the upper limit for the abundance of solid NH$_3$ relative to water ice is $<$5 $\%$ for ST6. 

The CH$_3$OH ice shows an absorption band at 9.75 $\mu$m (C--O stretching mode), which is not seen in the spectrum. 
The IRS spectrum shows weak emission lines of molecular hydrogen at 6.91 $\mu$m (0--0 S(5)), 9.67 $\mu$m (0--0 S(3)), 12.28 $\mu$m (0--0 S(2)), 17.04 $\mu$m (0--0 S(1)), 28.22 $\mu$m (0--0 S(0)). 
The 9.75 $\mu$m CH$_3$OH band is blended with a H$_2$ 0--0 S(3) line at 9.67 $\mu$m. 
Thus it is difficult to place an upper limit for the optical depth of the CH$_3$OH ice band. 

\begin{figure}[!]
\begin{center}
\includegraphics[width=9cm, angle=0]{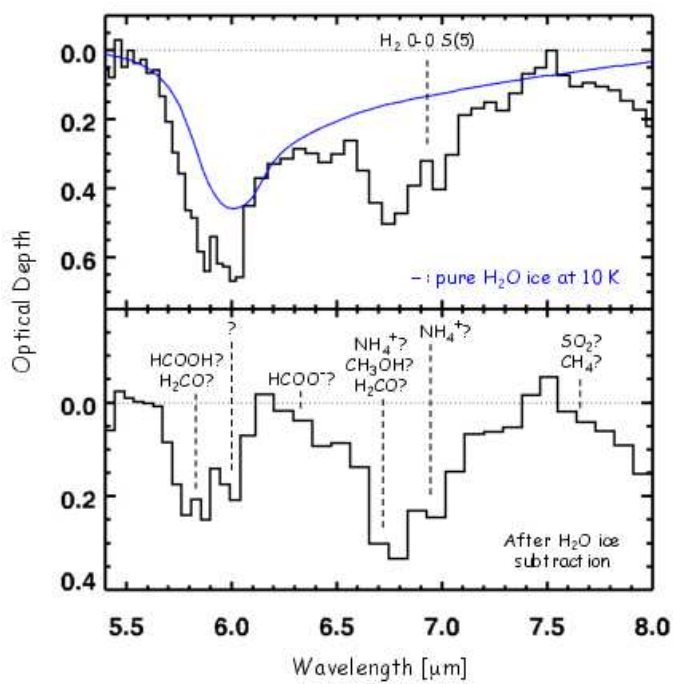}
\caption{
Upper panel: optical depth spectrum of ST6 in the 5.5--8 $\mu$m region. 
The smooth solid line (blue) indicates the expected spectrum of the 6.0 $\mu$m water ice band, whose strength is estimated based on the 3.05 $\mu$m band. 
Lower panel: spectrum after subtraction of the 6.0 $\mu$m water ice band. 
Absorption components that cannot be explained by the water ice are seen in the spectrum. 
Possible identifications of these absorption bands are indicated. 
}
\label{ST6_IRS2}
\end{center}
\end{figure}

Figure \ref{ST6_IRS2} shows the IRS spectrum in the 5--8 $\mu$m region. 
Several absorptions are seen in addition to the water ice absorption band at 6.0 $\mu$m (bending mode). 
To estimate the contribution of these additional absorption components, we estimate the strength of the 6.0 $\mu$m water ice band based on the column density derived from the 3.05 $\mu$m band and subtracted from the spectrum. 
We assume the laboratory spectrum of pure water ice at 10 K (data taken from the Leiden database) as a profile of the 6.0 $\mu$m band. 
The residual spectrum is shown in the lower panel of the figure together with possible molecules that contribute to the absorption bands \citep{Boo08,Boo15}. 
The figure hints at the possibility that various complex species could exist in the low metallicity environment of the LMC. 
Higher spectral- and spatial-resolution observations will enable us a more detailed analysis of these minor solids bands for a larger number of LMC and SMC YSOs.

%%%%%%%%%%
%%%%%%%%%%

\section{Discussions} 
The present data enable us to discuss spectral properties of the C--H stretching vibration region of embedded high-mass YSOs in the LMC. 
In this section, we discuss methanol ice chemistry and possible carriers of the 3.47 $\mu$m band in the LMC. 
Implications for the formation of organic molecules in low metallicity environments are also discussed.

\subsection{Methanol ice in the LMC} 
\subsubsection{Comparison of methanol ice abundance with Galactic sources} 
\begin{figure}[!]
\begin{center}
\includegraphics[width=9.1cm, angle=0]{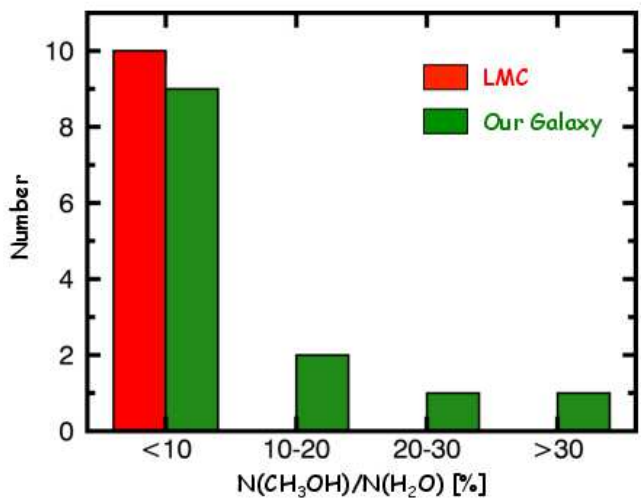}
\caption{
Histogram of the CH$_3$OH ice abundances for the LMC high-mass YSOs (red) and Galactic high-mass YSOs (green). 
All the LMC samples show the CH$_3$OH ice abundance less than 10 $\%$ relative to water ice, while several Galactic samples show higher CH$_3$OH ice abundances. 
}
\label{histo_CH3OH}
\end{center}
\end{figure}

Figure \ref{histo_CH3OH} compares the abundance of the CH$_3$OH ice between LMC and Galactic high-mass YSOs. 
An abundance of ice is defined as a ratio of a column density relative to the water ice column density, since water is the most abundant solid species in dense clouds. 
The plotted data in the figure are summarized in Tables \ref{Tab_column} and \ref{Tab_MW2}. 
For all the lines of sight of the present ten LMC YSOs, the CH$_3$OH ice abundance is less than 5--8 $\%$. 
Even for the sources with CH$_3$OH ice detection (ST6 and ST10), the derived abundances are as small as 5.6 and 3.7 $\%$, respectively. 
On the other hand, four out of 13 (about one-third of the total) Galactic samples show  CH$_3$OH ice abundances between 10 $\%$ and 40 $\%$. 
Although statistical uncertainties still remain as a result of the small number of samples, the present results suggest that solid CH$_3$OH is less abundant in the LMC than in our Galaxy within the present investigations. 
\citet{Sea10} also suggested  low CH$_3$OH ice abundance in LMC YSOs on the basis of the spectral analysis of the 15.2 $\mu$m CO$_2$ ice band. 
However, determination of CH$_3$OH ice abundances based on the profile analysis of the CO$_2$ ice band entails various uncertainties, as the authors cautioned in their paper. 
The present analysis provides direct evidence of the low CH$_3$OH ice abundance in LMC YSOs based on the measurement of the C--H stretching vibration band of the CH$_3$OH ice. 

\begin{figure}[!]
\begin{center}
\includegraphics[width=8.5cm, angle=0]{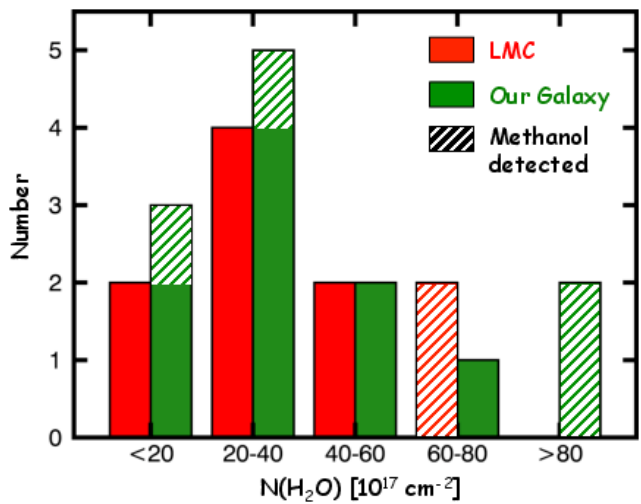}
\caption{
Histogram of the H$_2$O ice column densities for the LMC (red) and Galactic (green) high-mass YSOs. 
The number of samples for which the CH$_3$OH ice absorption is detected is indicated with transverse lines. 
}
\label{histo_H2O}
\end{center}
\end{figure}

Figure \ref{histo_H2O} shows a histogram of H$_2$O ice column densities for the LMC and Galactic samples. 
Since water is the most abundant ice species except molecular hydrogen, its column densities dominate the total ice column density along the line of sight. 
Although \citet{Oli11} suggested that the H$_2$O ice is selectively depleted in LMC YSOs, the present figure shows that the distribution of H$_2$O ice column densities is similar between the observed LMC YSOs and Galactic samples except for the extremely embedded sources, which are discussed below. 
Further, luminosities and appearances of infrared spectra are also similar between the LMC and Galactic samples \citep[][see also Fig. \ref{Spec}--\ref{Spec} in this work]{ST10,Gib04}. 
These facts suggest that the present LMC samples share similar properties (e.g., evolutionary stages) with the Galactic samples. 
Nevertheless, we did not detect any source with a high CH$_3$OH ice abundance, which indicates that the solid methanol is less abundant in the LMC. 
However, in the present LMC samples, we are still missing extremely embedded sources (N(H$_2$O) $>$ 10$^{19}$ cm$^{-2}$), such as W33A or AFGL7009S in our Galaxy. 
Although such extreme sources are not yet detected in the LMC, their infrared spectral information will be crucial to improve our understanding of ice chemistry in low metallicity environments. 
Further systematic observations of embedded YSOs in the LMC are obviously important.

%%%%%%%%%%
%%%%%%%%%%

%%%%%%%%%%
%%%%%%%%%%

%%%%%%%%%%%%%%%%%%%%%%%%%%%\begin{landscape}
\begin{table*}
\centering
\caption{Ice column densities and abundances for Galactic high-mass YSOs}
\label{Tab_MW2}
\begin{tabular}{ l c c c c c c c}
\hline\hline
Object               &  \textit{N}(H$_2$O)           &  \textit{N}(CH$_3$OH)\tablefootmark{*}   &  \textit{N}(CO)              &  \textit{N}(CO$_2$)       &  \textit{N}(CH$_3$OH)         &  \textit{N}(CO)                     &  \textit{N}(CO$_2$)                      \\
  &   &   &   &  &  /\textit{N}(H$_2$O)  & /\textit{N}(H$_2$O)  & /\textit{N}(H$_2$O)   \\
                         &  (10$^{17}$ cm$^{-2}$)      &  (10$^{17}$ cm$^{-2}$)                         &  (10$^{17}$ cm$^{-2}$)   &  (10$^{17}$ cm$^{-2}$)  &  ($\%$)     &   ($\%$)  &  ($\%$)     \\
\hline  
S140 IRS 1       &  19\tablefootmark{a}          &  $<$0.9\tablefootmark{b}                       &  ...                            &  4.2\tablefootmark{c}          &  $<$4.5                              &  ...                  &    22.1                          \\
Mon R2 IRS 2    &  35.6\tablefootmark{a}      &  1.9\tablefootmark{d}                             &  2.7\tablefootmark{a}  &  6.0\tablefootmark{c}          &  5.4                                   &  7.6                 &    16.9                          \\
Mon R2 IRS 3    &  19\tablefootmark{a}         &  $<$1.3\tablefootmark{b}                       &  ...                            &  1.6\tablefootmark{c}          &  $<$6.7                              &  ...                 &      8.4                          \\
RAFGL989        &  22.4\tablefootmark{e}       &  0.7\tablefootmark{e}                            &  4.5\tablefootmark{a}  &  8.1\tablefootmark{c}          &  3.3                                    &  20                &      36.2                          \\
RAFGL2136      &  45.7\tablefootmark{e}       & 3.5\tablefootmark{d}                             &  2.7\tablefootmark{a}  &  7.8\tablefootmark{c}          &  7.6                                    &  5.9                &     17.1                          \\
RAFGL2591      &  12\tablefootmark{a}          &  2.4\tablefootmark{a}                            &  ...                            &  1.6\tablefootmark{c}          &  20.3                                  &  ...                   &    13.3                           \\
RAFGL7009S    &  113.1\tablefootmark{e}     & 35.5\tablefootmark{b}                            &  18\tablefootmark{a}   &  25\tablefootmark{c,+}          &  31.4                                  &  16                 &      22.1                             \\
W33 A              &  125.7\tablefootmark{e}     &  19.9\tablefootmark{d}                           &  8.9\tablefootmark{a}  &  14.5\tablefootmark{c}          &  15.8                                  &  7.1               &     11.5                            \\
NGC 7538 IRS1 &  22\tablefootmark{a}         &  $<$1.1\tablefootmark{b}                       &  1.8\tablefootmark{a}  &  5.1\tablefootmark{c}          &  $<$5.1                               &  8.2                &     23.2                          \\
NGC 7538 IRS9 &  64.1\tablefootmark{e}      & 4.1\tablefootmark{d}                              &  12\tablefootmark{a}   &  16.3\tablefootmark{c}          &  6.4                                    &  18.7              &     25.4                            \\
Orion BN           &  25\tablefootmark{a}         &  $<$1.1\tablefootmark{b}                       &  ...                            &  2.9\tablefootmark{c}          &  $<$4.5                               &  ...                 &      11.6                         \\
Orion IRC2        &  24.5\tablefootmark{a}       &  3.6\tablefootmark{a}                            &  ...                            &  2.6\tablefootmark{c}          &  14.6                                   &  ...                  &     10.6                           \\
W3 IRS 5          &  56.5\tablefootmark{e}          &  $<$2.3\tablefootmark{b}                      &  1.6\tablefootmark{a}   &  7.1\tablefootmark{c}          &  $<$4.1                              &  3.1                 &     12.6                         \\
\hline
\end{tabular}
\tablefoot{All the values are taken from the literature. 
\tablefoottext{*}{Methanol column densities are recalculated using the band strength of 5.3$\times$10$^{-18}$ cm molecule$^{-1}$ and $\Delta$$\nu$ of 31 cm$^{-1}$, except for RAGL7009S, whose column density is estimated from the combination modes at 3.9 $\mu$m. }
\tablefoottext{+}{Estimated from the 15.2 $\mu$m absorption band. } \\
Ref. 
\tablefoottext{a}{\citet{Gib04}}; 
\tablefoottext{b}{\citet{Dar99}}; 
\tablefoottext{c}{\citet{Gib04}}; 
\tablefoottext{d}{\citet{Bro99}}; 
\tablefoottext{e}{\citet{Boo08}
}}
\end{table*}
%%%%%%%%%%%%%%%%%%%%%%%%%%

\subsubsection{Warm ice chemistry} 
We propose the warm ice chemistry hypothesis to explain the characteristic chemical properties of ices in the LMC revealed by the present and previous observations. 

It is widely accepted that CH$_3$OH is mainly formed by solid-phase reactions. 
One possible reaction mechanism is caused by diffusive grain surface chemistry. 
Laboratory experiments suggest that hydrogenation of CO plays a key role in the formation of CH$_3$OH \citep[e.g.,][]{Wat02,Wat07}. 
The temperature dependance of the formation of H$_2$CO and CH$_3$OH by hydrogenation is investigated by experiments, which have reported that their formation is significantly suppressed when the surface temperature is higher than 20 K. 
Numerical simulations of grain surface chemistry also suggest that the formation efficiency of CH$_3$OH decreases as the temperature of dust grains increase \citep[e.g., ][]{Ruf01,Cup09,Cha12}. 
These studies suggest that hydrogenation of CO, which leads to the formation of CH$_3$OH, becomes less efficient at high dust temperatures. 
These behaviors can be explained qualitatively by very rapid diffusion or evaporation of hydrogen atoms at increased surface temperature. 

The binding energy of a hydrogen atom on the ice surface is known to be much smaller than those of other species \citep[][and references therein]{HW13}. 
When dust temperature is sufficiently low ($\sim$10 K), hydrogen atoms can diffuse on the surface at a moderate speed to find and to react with CO. 
At increased temperatures ($\sim$20 K), hydrogen starts to diffuse very rapidly on the surface, which suppresses hydrogenation because hydrogen atoms move to other surface sites before reacting with CO even though they could meet on the surface. 
At even higher temperatures, hydrogen atoms start to evaporate rapidly from the surface, which further suppresses hydrogenation. 

A detailed study of solid CH$_3$OH for Galactic YSOs and quiescent clouds suggest that the efficiency of CH$_3$OH production in dense cores and protostellar envelopes is mediated by the degree of prior CO depletion \citep{Whi11}. 
Abundances of solid CO in the lines of sight of the present LMC high-mass YSOs are summarized in Table \ref{Tab_column}, while those for Galactic sources are in Table \ref{Tab_MW2}. 
Mean CO ice abundances and standard deviations excluding sources with upper limit are 9.4 $\pm$ 5.3 for the LMC samples and 10.8 $\pm$ 6.5 for the Galactic samples. 
The similar CO ice abundance for LMC and Galactic sources suggests that freeze out of CO gas onto grain surfaces occurs even in the LMC. 
Sufficient freeze out of CO is also suggested previously for several LMC YSOs \citep{Oli11}. 
Therefore, we suggest that less efficient hydrogenation of CO due to warm dust temperatures suppress production of solid CH$_3$OH around high-mass YSOs in the LMC rather than inefficient depletion of CO. 

An alternative formation mechanism for solid CH$_3$OH involves energetic processes of ice mantles, such as the formation through UV photon irradiation (photolysis) or proton irradiation (radiolysis), which are proposed by experimental studies \citep[e.g., ][]{Hud99,Ger01,Bar02,Wat07}. 
The UV radiation field in dense shielded regions of molecular clouds is dominated by cosmic-ray induced UV photons. 
The strength of cosmic-ray induced UV radiation field is closely related to the cosmic-ray density, which is reported to be lower in the LMC than in our Galaxy \citep{Abd10}. 
This indicates that energetic processes induced by cosmic rays are expected to be less efficient in dense clouds in the LMC. 
Thus, if the formation of solid CH$_3$OH is dominated by energetic processes, the low cosmic-ray density in the LMC can account for the observed low CH$_3$OH ice abundance. 
However, this does not reject the hypothesis of warm ice chemistry because the reduction of energetic processes cannot explain the increased abundance of the CO$_2$ ice, which is discussed in Section 4.1.3. 
With the current observational data, it is difficult to separate the relative contribution of warm ice chemistry and the low cosmic-ray density on the suppressed formation of CH$_3$OH in the LMC. 
On the other hand, the strong interstellar UV radiation field in the LMC does not contribute to the energetic formation processes of CH$_3$OH. 
This is because the formation of the CH$_3$OH ice requires much denser and more shielded environments (Av $\sim$17) than those required for the H$_2$O ice formation (Av $\sim$6), as discussed in \citet{Whi11}. 
The interstellar UV photons are heavily attenuated in such very dense clouds, and the CH$_3$OH ice formation is not  affected by the strength of the external UV radiation field. 

Water ice, which is abundantly detected in the LMC, is also mainly formed by hydrogenation of surface species in dense clouds. 
One possible reaction pathway for water ice formation is by $OH + H \to H_2O$ \citep[e.g., ][and references therein]{vanD13}. 
Under the warm ice chemistry hypothesis, it should be explained why hydrogenation of OH can occur in the LMC, while hydrogenation of CO is suppressed. 
We speculate that this is caused by the difference of an activation barrier in each hydrogenation pathway. 
It is known that the surface reaction of $CO + H \to HCO$ and $H_2CO + H \to H_3CO$ possess a high activation barrier \citep[$>$ 400 K, ][]{Woo02,Fuc09}. 
This indicates that CO and H have to stay on the same or adjacent surface site for a certain amount of time to react. 
In this case, hydrogenation is suppressed at high temperatures because of very rapid diffusion or evaporation as mentioned above. 
On the other hand, the hydrogenation of OH to H$_2$O is known to be barrierless. 
Very rapid diffusion of hydrogen does not suppress hydrogenation of OH because their reaction proceeds immediately after the encounter of reactants. 
In addition, the Eley-Rideal mechanism, which is caused by direct collisions of the gas-phase species to the surface species, may also contribute to the formation. 
Thus, we speculate that formation of water ice is still possible even at moderately warm dust temperatures where CO hydrogenation becomes less efficient. 

In the LMC, it is highly possible that the typical temperature of molecular clouds is higher than in our Galaxy due to stronger interstellar radiation field. 
The temperature of dust in molecular clouds should be closely related to metallicity of the galaxy since the interstellar radiation field is attenuated by dust grains. 
Several studies actually argue that the typical dust temperature of diffuse clouds in the LMC is higher than in our Galaxy based on far-infrared to sub-millimeter observations \citep{Agu03,Sak06}. 
\citet{vanL10_b} reported that dust around high-mass YSOs and compact HII regions in the SMC is warmer (37--51 K) than those in comparable objects in the LMC (32-44 K). 
For further extragalaxies, it is reported that the color temperature of galaxies measured by far-infrared observations is negatively correlated with their metallicities \citep[e.g.,][]{Eng08}. 
Although dust temperatures of {dense} clouds in the LMC, where significant amount of ices are formed, are still an open question, the above studies imply the higher dust temperatures of dust in lower metallicity environments. 

Spectral profiles of the 15.2 $\mu$m CO$_2$ ice band for high-mass YSOs in the LMC are well studied with \textit{Spitzer} observations \citep{Oli09,Sea10}. 
\citet{Sea10} suggest that the CO$_2$ ices around high-mass YSOs in the LMC are more thermally processed than those in our Galaxy. 
This also supports the idea that the increased dust temperature in the LMC affects the properties of ices around embedded YSOs. 

The warm ice chemistry scenario suggests that the low CH$_3$OH ice abundance in the LMC favor warmer dust temperatures in dense shielded regions where ices are formed. 
Future measurements of dust and gas temperatures of compact dense clouds in the LMC will be crucial for testing the present hypothesis and for understanding of ice chemistry in low metallicity environments.

\subsubsection{Decrease of CH$_3$OH and increase of CO$_2$ as a consequence of warm ice chemistry}
\begin{figure}[!]
\begin{center}
\includegraphics[width=9.2cm, angle=0]{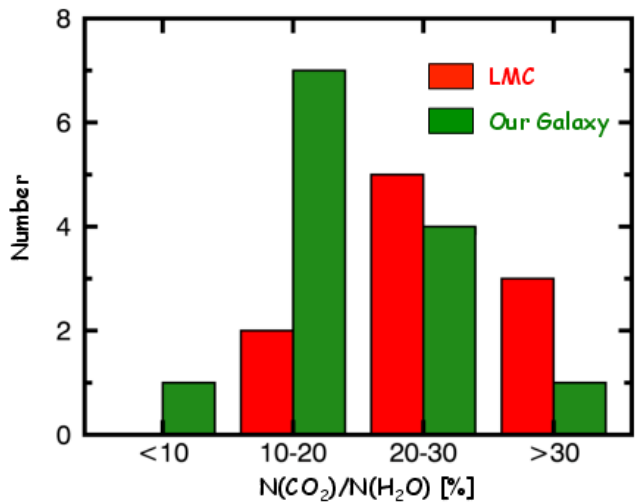}
\caption{
Histogram of the CO$_2$ ice abundances for the LMC high-mass YSOs (red) and Galactic high-mass YSOs (green). 
The LMC samples show the relatively high CO$_2$/H$_2$O ice ratio compared to the Galactic samples. 
}
\label{histo_CO2}
\end{center}
\end{figure}

The relatively high abundance of the CO$_2$ ice in the LMC, which was reported in previous infrared observations \citep{ST,ST10,Oli09,Sea10}, is also consistent with the above warm ice chemistry scenario. 
The CO$_2$/H$_2$O ice ratio for the present LMC high-mass YSOs are compared with those of Galactic high-mass YSOs in Fig. \ref{histo_CO2}. 
The mean value and standard deviation of the CO$_2$/H$_2$O ice ratio is 28.5 $\pm$ 11.6 $\%$ for the ten LMC samples and 17.8 $\pm$ 7.8 $\%$ for the 13 Galactic samples, respectively (see Tables \ref{Tab_column} and \ref{Tab_MW2}). 
The median value of the CO$_2$/H$_2$O ice ratio is 29.1 $\%$ for the LMC samples and 16.9 $\%$ for the Galactic samples, respectively. 
The high-mass YSOs in the LMC show the higher CO$_2$ ice abundance compared to Galactic high-mass YSOs as suggested in previous studies. 

It is known that the reaction of $CO + OH \to CO_2 + H$ is one of the dominant pathways for the formation of the CO$_2$ ice \citep[e.g.,][]{Oba10,Iop11}. 
As suggested by numerical simulations \citep[e.g.,][]{Ruf01,Cha12}, this formation pathway becomes efficient as the dust temperature increases because the mobility of CO on the surface increases accordingly. 
Since the binding energy of CO on the ice surface is much larger than that of hydrogen, CO can still remain and moderately diffuse on the surface even at high dust temperatures at which hydrogen atoms evaporate. 
Therefore, the high dust temperature in the LMC can enhance the formation of CO$_2$ by increasing the mobility of CO, while it can reduce the formation of CH$_3$OH by suppressing the hydrogenation of CO. 
The lower elemental C/O ratio in the LMC than in our Galaxy (see $\S$4.2.2) can also contribute to the low abundance of CH$_3$OH relative to H$_2$O. 
However,  we emphasize that the CO$_2$ ice abundance is enhanced in the LMC despite the low C/O ratio. 
Warm ice chemistry can simultaneously account for the enhanced formation of CO$_2$ and the suppressed formation of CH$_3$OH. 
This would suggest that warm ice chemistry is responsible for the characteristic chemical compositions of ices in the LMC. 

\citet{Oli11} argued that the high CO$_2$/H$_2$O ice ratio in the LMC is caused by the depletion of the H$_2$O ice rather than the enhanced formation of the CO$_2$ ice. 
However, the enhanced formation of CO$_2$ is more consistent with the observed low abundance of the CH$_3$OH ice in the LMC, since both are expected in grain surface chemistry at a relatively high dust temperature, as discussed above. 
The H$_2$O ice column densities are similar between the present LMC samples and Galactic samples except for two extremely embedded Galactic sources, as discussed in $\S$4.1.1. 
Therefore, we suggest that enhanced formation of CO$_2$ is responsible for the high CO$_2$/H$_2$O ice ratio in the LMC.

Based on the above discussions, we conclude that warm ice chemistry is responsible for the characteristic chemical compositions of ices around high-mass YSOs in the LMC. 
The present results suggest that conditions to form CH$_3$OH are less often encountered in the LMC's environment presumably due to warm ice chemistry. 
A key factor for warm ice chemistry is less efficient hydrogenation of surface species, which is triggered by the elevated temperature of dust grains in dense molecular clouds. 
Dust temperatures of ice-forming dense clouds in metal-poor environments are expected to be higher than those in metal-rich environments, as discussed in the previous section.  
We therefore suggest that warm ice chemistry is one of the characteristics of interstellar and circumstellar chemistry of dense ISM in low metallicity galaxies.

%%%%%%%%%%
%%%%%%%%%%

\subsubsection{Gas-phase methanol in the LMC}
Previous radio observations have reported the underabundance of methanol masers in the LMC in comparison to our Galaxy \citep{Bea96,Gre08,Ell10}. 
Thermal emission lines of methanol are detected toward a limited number of HII regions in the LMC \citep{Hei99,Wan09}. 
Recent unbiased spectral line surveys toward the LMC's molecular clouds suggest that thermal emission lines due to gas-phase methanol is significantly weaker in the LMC than in our Galaxy \citep[Shimonishi et al. in prep.;][]{Nis15}. 
We suggest that the low production rate of solid methanol due to warm ice chemistry contributes to few detections of methanol gas in the LMC.

\subsection{The 3.47 $\mu$m band in the LMC} 
In contrast to the very weak or absent CH$_3$OH ice absorption in the LMC, six out of eleven sources show the 3.47 $\mu$m absorption band. 
We here discuss the difference and similarity of the band between the LMC and our Galaxy to understand properties of the band carrier in low metallicity environment.

\subsubsection{Possible carriers of the 3.47 $\mu$m band} 
The 3.47 $\mu$m absorption band is widely detected toward a variety of embedded sources such as high- to intermediate-mass YSOs \citep[e.g., ][]{All92,Bro96,Bro99,Dar01,Dar02,Ish02}, low-mass YSOs \citep[e.g., ][]{Pon03,Thi06}, and quiescent dense molecular clouds \citep[e.g., ][]{Chi96,Chi11}. 
Despite many spectroscopic detections, the carrier of the 3.47 $\mu$m band is still under debate and various candidates have been proposed in the literature. 

\citet{All92} argued that the C--H vibration of hydrogen atoms bonded to tertiary carbon atoms could be responsible for the 3.47 $\mu$m band 
\footnote{There are three classes of carbon atoms related to C--H vibrations of aliphatic hydrocarbons; primary, secondary, and tertiary aliphatic carbons. 
In the primary aliphatic carbon,  three of its four single bonds are directed to other carbons. 
Similarly, in the secondary and the tertiary carbon,  one or two of its four bonds are directed to other carbons \citep{All92}. }. 
Although the primary and secondary hydrocarbons also show characteristic bands in 3.4--3.5 $\mu$m regions \citep[e.g., ][]{Men10}, the predominance of the 3.47 $\mu$m band in various embedded sources in our Galaxy puts constraints on the structure of hydrocarbons in dense clouds. 
The authors suggest that the only way in which tertiary carbon atoms can dominate the spectrum is if the carbon atoms are arranged in a diamond structure. 
These interstellar diamonds may have a connection with nano-sized diamonds in meteorites \citep[e.g., ][]{Lew87,Pir07}. 

Alternatively, \citet{Dar01} and \citet{Dar02} argued that the 3.47 $\mu$m band is related to an ammonia hydrate formed in NH$_3$:H$_2$O mixture ice. 
The authors show that a stretching vibration mode arising from the interaction between a nitrogen atom of NH$_3$ and an O--H bond of H$_2$O can account for a part of the 3.47 $\mu$m absorption. 
It was previously reported for Galactic sources that optical depths of the 3.47 $\mu$m band correlate well with those of the H$_2$O ice band \citep{Bro99}. 
An ammonia hydrate, as a carrier of the 3.47 $\mu$m band, is thus consistent with the correlation of the 3.47 $\mu$m band and the H$_2$O ice.

%%%%%%%%%%
\begin{figure*}[!]
\begin{center}
\includegraphics[width=18cm, angle=0]{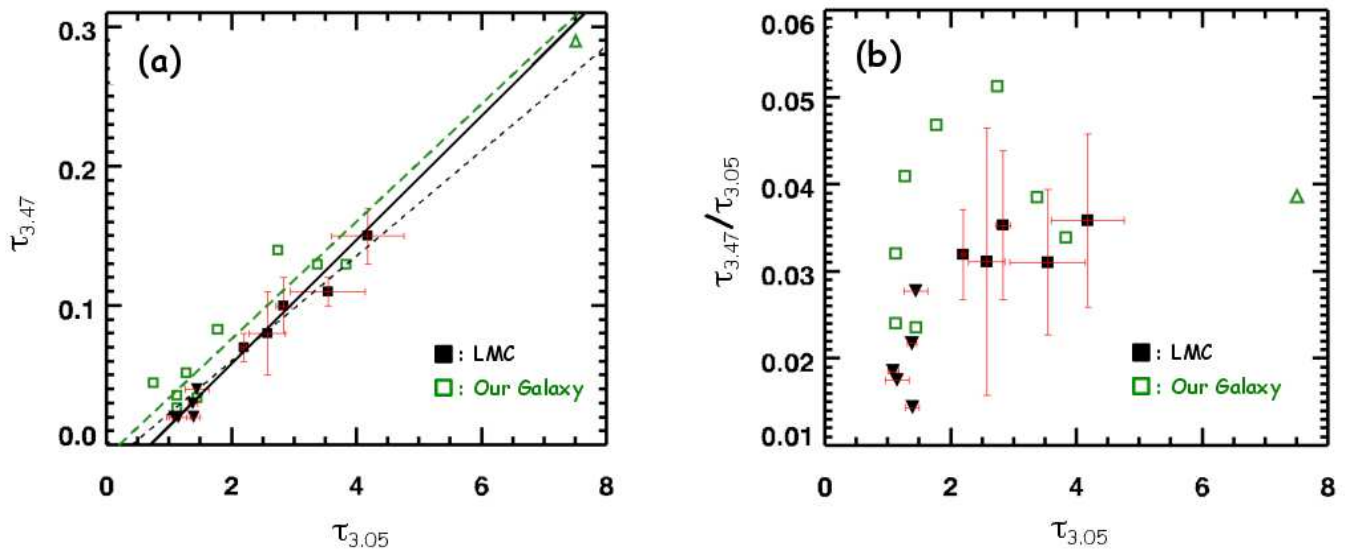}
\caption{
(a) Optical depths of the 3.05 $\mu$m H$_2$O ice band vs. the 3.47 $\mu$m band. 
Filled (black) and open (green) squares represent data points of the LMC high-mass YSOs (this work) and Galactic high-mass YSOs (see Table \ref{Tab_MW} for references). 
The upward and downward triangles represents the lower and upper limit. 
Solid (black) and dashed (green) lines represent the results of a straight-line fit to LMC and Galactic data points, respectively. 
The best-fit line is $\tau_{3.47} = 0.044\tau_{3.05} - 0.030$ for the LMC, and  $\tau_{3.47} = 0.043\tau_{3.05} - 0.008$ for our Galaxy. 
The straight-line fit without upper limit points is also shown with black dashed line for the LMC data. 
The figure shows that the slope of the $\tau_{3.47}$ and $\tau_{3.05}$ correlation is similar between the LMC and Galactic sources. 
(b) Optical depth ratio of $\tau_{3.47}$/$\tau_{3.05}$ vs. $\tau_{3.05}$. 
The symbols are same as in the left panel. 
For the LMC sources with $\tau_{3.05}$ $<$ 2, the $\tau_{3.47}$/$\tau_{3.05}$ ratio is significantly lower than Galactic sources, which suggests the presence of threshold $\tau_{3.05}$ for the appearance of the 3.47 $\mu$m band in the LMC. 
}
\label{t31_t347}
\end{center}
\end{figure*}
%%%%%%%%%%

%%%%%%%%%%%%%%%%%%%%%%%%%%%\begin{landscape}
\begin{table}[b]
\centering
\caption{Optical depths of the water ice and the 3.47 $\mu$m absorption bands for Galactic high-mass YSOs}
\label{Tab_MW}
\begin{tabular}{ l c c}
\hline\hline
Object                  & $\tau$$_{3.05 \mu m}$  & $\tau$$_{3.47 \mu m}$    \\
\hline  
S140 IRS 1         & 1.12\tablefootmark{a}     & 0.027\tablefootmark{b}       \\
Mon R2 IRS 2     & 1.77\tablefootmark{a}      & 0.083\tablefootmark{c}     \\
Mon R2 IRS 3     & 1.12\tablefootmark{a}      & 0.036\tablefootmark{c}      \\
RAFGL989         & 1.34\tablefootmark{+}      & ...                                 \\
RAFGL2136       & 2.73\tablefootmark{+}      & 0.14\tablefootmark{c}      \\
RAFGL2591       & 0.74\tablefootmark{a}      & 0.045\tablefootmark{c}       \\
RAFGL7009S     & 6.75\tablefootmark{+}     & ...                                    \\
W33 A                 & 7.50\tablefootmark{+}      & $>$0.29\tablefootmark{c}     \\
NGC 7538 IRS1 & 1.27\tablefootmark{a}      & 0.052\tablefootmark{b}      \\
NGC 7538 IRS9 & 3.83\tablefootmark{+}      & 0.13\tablefootmark{c}      \\
Orion BN            & 1.44\tablefootmark{a}      & 0.034\tablefootmark{b}     \\
Orion IRC2         & 1.48\tablefootmark{a}      & ...                                   \\
W3 IRS 5           & 3.37\tablefootmark{+}      & 0.13\tablefootmark{b}         \\
\hline
\end{tabular}
\tablefoot{
\tablefoottext{+}{Optical depths of the H$_2$O ice are calculated based on the corresponding column densities in Table \ref{Tab_MW2}  assuming $\Delta$$\nu$ = 335 cm$^{-1}$ and A = 2.0$\times$10$^{-16}$ cm molecule$^{-1}$. }
\\
%{\bf References. }
Ref. 
\tablefoottext{a}{\citet{Gib04}}; 
\tablefoottext{b}{\citet{Bro96}}; 
\tablefoottext{c}{\citet{Bro99}}; 
}
\end{table}
%%%%%%%%%%%%%%%%%%%%%%%%%%

\subsubsection{Comparison of the band properties with Galactic sources} 
Figure \ref{t31_t347}a compares the optical depths of the 3.47 $\mu$m band and the 3.05 $\mu$m H$_2$O ice band for the LMC and Galactic samples. 
The plotted data are summarized in Tables \ref{Tab_tau} and \ref{Tab_MW}. 
The above-mentioned correlation of the 3.47 $\mu$m band and the H$_2$O ice observed for Galactic sources is also seen in the LMC sources, as shown in the figure. 
A possible explanation for the observed correlation is that the carrier of the 3.47 $\mu$m band is formed by a process similar  to that of water ice formation. 
This process is presumably related to hydrogenation of carbon or nitrogen atoms on the dust surface since water ice is also formed by hydrogenation. 

The figure suggests that the 3.47 $\mu$m band and the H$_2$O ice band correlate similarly between the LMC and Galactic samples, but the LMC sources seem to require a higher H$_2$O ice threshold for the appearance of the 3.47 $\mu$m band. 
A least-squares fit of a straight line to the LMC data points gives
\begin{equation}
\tau_{3.47} = (0.044 \pm 0.004)\tau_{3.05} - (0.030 \pm 0.007), 
\end{equation}
where the fitting is not weighted and upper limits are included as data point. 
When the upper limits are excluded in the fit, the slope is 0.038 $\pm$ 0.011 and the intercept is 0.015 $\pm$ 0.031. 
This line, however, does not account for the upper limit points as seen in the figure. 
The same fitting for the Galactic data points gives 
\begin{equation}
\tau_{3.47} = (0.043 \pm 0.004)\tau_{3.05} - (0.008 \pm 0.005) 
.\end{equation}
An intercept value of $\tau_{3.05}$ at $\tau_{3.47}$ = 0 axis (i.e., threshold $\tau_{3.05}$) is slightly different between the LMC and our Galaxy. 
The threshold value is estimated to be $\tau_{3.05}$ = 0.69 $\pm$ 0.23 for the LMC and $\tau_{3.05}$ = 0.19 $\pm$ 0.14 for our Galaxy, respectively. 
The difference may suggest that a more shielded environment is necessary for the formation of the 3.47 $\mu$m band carrier in the LMC. 
Figure \ref{t31_t347}b compares the $\tau_{3.47}$/$\tau_{3.05}$ ratio against $\tau_{3.05}$. 
The LMC sources with $\tau_{3.05}$ $<$ 2 show significantly lower $\tau_{3.47}$/$\tau_{3.05}$ ratio than those of Galactic sources. 
This again suggests the non-zero value of threshold $\tau_{3.05}$ for the presence of the 3.47 micron absorption in the LMC. 
However, the ratio may not be the best representation once it is seen that there is a threshold. 
Follow-up observations toward a larger number of embedded sources in the LMC are highly required to provide solid evidence on the observed difference of the threshold $\tau_{3.05}$ in the LMC.

On the other hand, the comparison of the samples in which the 3.47 $\mu$m band is detected suggests that the ratio of $\tau_{3.47}$ and $\tau_{3.05}$ is not significantly different between the LMC and our Galaxy. 
The mean value and standard deviation of $\tau_{3.47}$/$\tau_{3.05}$ ratio are 0.0331 $\pm$ 0.0024 for the five LMC samples and 0.0392 $\pm$ 0.0124 for the nine Galactic samples, respectively. 
The median value of $\tau_{3.47}$/$\tau_{3.05}$ ratio is 0.0320 for the LMC samples and 0.0386 for the Galactic samples, respectively. 
The LMC samples show only slightly lower $\tau_{3.47}$/$\tau_{3.05}$ ratios compared to Galactic samples (see also Figure \ref{t31_t347}b). 
This would suggest that, in the well-shielded regions where the 3.47 $\mu$m band is detected, the lower metallicity as well as the different elemental abundances and different interstellar environment of the LMC have little effect on the abundance ratio of the 3.47 $\mu$m band carrier and water ice. 

In the LMC, the elemental abundance ratio of both C/O and N/O are lower than those in the sun: [C/O]$_{LMC}$/[C/O]$_{Sun}$ $\sim$0.55, [N/O]$_{LMC}$/[N/O]$_{Sun}$ $\sim$0.30 \citep{Duf82}. 
An underabundance of the gas-phase atomic nitrogen in the ISM of the LMC is also confirmed by \textit{Spitzer}/MIPS far-infrared observations \citep{vanL10}. 
Hence, abundances of both carbon bearing and nitrogen bearing species relative to oxygen bearing species decrease if their ratios simply depend on the initial elemental abundances. 
However, formation efficiencies of interstellar molecules are not that straightforward, and   previous studies actually reported that the CO$_2$/H$_2$O ice ratio is higher in the LMC than in our Galaxy despite the lower C/O ratio in the LMC \citep[e.g.,][]{ST10}. 
This fact suggests that the simple argument from the low C/O ratio cannot rule out C- or N-bearing species as a potential candidate of the 3.47 $\mu$m band.
In the next section, we discuss about possible carriers of the 3.47 $\mu$m band in the environment of the LMC.

\subsubsection{The carrier of the 3.47 $\mu$m band in the LMC}
The better correlation between $\tau_{3.47}$ and $\tau_{3.05}$ than the correlation of $\tau_{3.47}$ and $\tau_{9.7}$ (silicate band) reported in previous studies \citep{Bro96,Bro99} suggests that the formation of the 3.47 $\mu$m band carrier is related to dense and cold regions where ices are formed. 
As mentioned earlier, there are two hypotheses for the carrier of the 3.47 $\mu$m band, i.e., C--H bonds on diamonds and an ammonia hydrate. 
We discuss which hypothesis is more consistent with the interstellar environment and the observed properties of the 3.47 $\mu$m band in the LMC. 

Under the hypothesis that interstellar diamonds are responsible for the 3.47 $\mu$m band, the formation and destruction of C--H bonds on carbon grains should be considered. 
\citet{Men08,Men10} confirms the formation of tertiary C--H bonds after irradiation of hydrogen atoms to carbon particle samples coated by water ice. 
In this case, hydrogenation of carbon atoms occurs at the interface of dust and ice by penetration of hydrogen atoms from the ice surface to the grain surface. 
Owing to warm ice chemistry in the LMC, as discussed in Section 4.1.2, hydrogenation of surface species is less efficient if there is an activation barrier because hydrogen atoms rapidly diffuse or evaporate on/from the surface before a reaction. 
This is caused by a relatively high dust temperature in LMC molecular clouds. 
These characteristics, however, should not be simply applied to the hydrogenation that proceeds at the interface of dust and ice. 
Currently, it is difficult to quantitatively discuss the temperature dependence of such an internal hydrogenation reaction because of a lack of relevant laboratory information, and thus the formation efficiency of tertiary C--H bonds in the LMC remains to be explained. 

On the other hand, C--H bonds in dense clouds are mainly destroyed by cosmic-ray hits and internal UV induced by interaction of cosmic rays and molecular hydrogen \citep{Men10}. 
The flux of cosmic-ray induced UV photons is proportional to cosmic-ray fluxes \citep[e.g., ][]{Pra83,She04}. 
It is reported based on gamma-ray observations that the cosmic-ray density in the LMC is 20--30$\%$ lower than in our Galaxy \citep{Abd10}. 
Such a low cosmic-ray density can decrease the destruction efficiency of C--H bonds in the LMC. 
If the 3.47 $\mu$m band dominantly arises from the C--H vibration of hydrogen atoms bonded to tertiary carbons, this could be a possible cause for the similar $\tau_{3.47}$/$\tau_{3.05}$ ratio between the LMC and our Galaxy despite the low C/O ratio in the LMC. 

Under the hypothesis that an ammonia hydrate is responsible for the 3.47 $\mu$m band, molecular abundance of ammonia in the LMC should be considered. 
Gas-phase ammonia is reported to be deficient in star-forming regions in the LMC; the fractional abundance is estimated to be 1.5--5 orders of magnitude lower than those in Galactic star-forming regions \citep{Ott10}. 
Recent spectral line surveys toward molecular clouds in the LMC suggest that nitrogen-containing gas-phase molecules, such as HCN, HNC, N$_2$H$^+$, are also underabundant in the LMC \citep{Nis15}. 
These studies suggest that the low elemental abundance of nitrogen (see $\S$4.2.2) as well the intense interstellar UV radiation field in the LMC contribute to the low abundances of nitrogen-bearing molecules. 

An upper limit is estimated for the abundance of solid ammonia in ST6, which shows the deepest 3.47 $\mu$m band among the sources we investigated. 
 The estimated upper limit of the solid ammonia abundance (NH$_3$/H$_2$O) is $<$5 $\%$ for ST6 based on the spectral analysis of the  \textit{Spitzer}/IRS data (see $\S$3.3). 
The abundance of solid ammonia, estimated on the basis of the ammonia hydrate hypothesis of the 3.47 $\mu$m band, is reported to be less than or equal to 7 $\%$ for the two Galactic high-mass YSOs RAFGL 989 and RAFGL 2136 \citep{Dar02} and about 5 $\%$ for a larger Galactic sample \citep{Dar01}. 
The solid ammonia is also observed toward Galactic low-mass YSOs through the umbrella mode at 9.0 $\mu$m \citep{Bot10}. 
The typical abundance is around 5 $\%$ over 23 sources in their Table 2, excluding the two sources mentioned in the article as likely upper limits and the source EC82, whose silicate dust continuum is clearly observed in emission. 
Hence, the upper limit of the solid ammonia abundance for the LMC source ST6 is not significantly different from Galactic values. 

Numerical simulations of grain surface chemistry suggest that the abundance of solid ammonia decreases as the temperature of dust grains increases \citep[e.g., ][]{Sta04,Cha14}. 
This behavior is interpreted as a decrease in hydrogenation efficiency of nitrogen on the dust surface. 
This would suggest that warm ice chemistry discussed in $\S$4.1.2 can lower the formation efficiency of solid ammonia in the LMC. 

Although a current observational constraint on solid ammonia abundance in the LMC is still less conclusive because of a sample, we speculate that the interstellar environment of the LMC favors the low abundance of solid ammonia, which is consistent with radio observations of gas-phase ammonia and numerical simulation of grain surface chemistry mentioned above. 
These discussions indicate that the contribution from ammonia hydrate may be less significant in the LMC. 
Our interpretation cannot exclude the contribution of an ammonia hydrate to the 3.47 $\mu$m band observed in our Galaxy. 
Further observational constraints on abundances of ammonia for larger number of LMC samples are key to improving our understanding of the 3.47 $\mu$m band carrier in the LMC.

\subsection{Implications for the formation of organic molecules in low metallicity galaxies} 
Complex organic molecules are widely detected toward Galactic chemically-rich sources such as hot cores and hot corinos \citep[e.g.,][]{vDB98,Her09}. 
Understanding of the formation and evolution of these molecules in space is of interest in exploring the chemical complexity of the interstellar medium. 
Particularly, properties of complex organic molecules in low metallicity environments are of great interest in order to discuss the building blocks of prebiotic molecules in the past metal-poor universe. 

Formation processes of complex organic molecules in interstellar and circumstellar environments are not well understood yet, but grain surface chemistry is believed to play an important role. 
Theoretical models suggest that CH$_3$OH is a key to the formation of complex organic molecules \citep[e.g.,][]{Mil97,NM04,Gar08a,Cha14}. 
The sublimation of solid CH$_3$OH into gas-phase and subsequent reactions in warm and dense circumstellar regions can enhance the formation of complex molecules \citep[e.g.,][]{NM04}. 
Alternatively, CH$_3$OH can produce heavy radicals like CH$_3$O or CH$_2$OH by photodissociation, which subsequently evolve into complex molecules by grain surface reactions \citep[e.g.,][]{Gar08a}. 
Thus, the low CH$_3$OH abundance in the LMC implies that formation of complex organic molecules from methanol-derived species is less efficient in the LMC. 
However, it is obvious that we should take  diverse mechanisms into account to discuss complex molecular chemistry. 
A variety of carbon-/oxygen-/nitrogen-bearing species (e.g., hydrocarbon, PAH) are possible building blocks of complex organic molecules. 
Further laboratory and theoretical studies on the formation of complex organic molecules are needed in conjunction with observational efforts to detect complex species in low metallicity galaxies.

\section{Summary}
We present the results of near-infrared spectroscopic observations toward embedded high-mass YSOs in the LMC using VLT/ISAAC. 
The medium-resolution ($R$ $\sim$500) spectra in the 3--4 $\mu$m region are presented for eleven sources in the LMC. 
The achieved spatial resolution is about 0.15 pc at the distance of the LMC, which is higher than those of previous satellite observations with \textit{AKARI} or \textit{Spitzer} by a factor of $\sim$10. 
The properties of detected ice absorption bands (water, methanol, 3.47 $\mu$m band) are investigated and we obtained the following conclusions. 

\begin{enumerate}
\item
The H$_2$O ice absorption band at 3.05 $\mu$m is detected for all of the YSO samples. 
Thanks to the high spatial resolution achieved by the VLT, contamination by PAH emission bands is significantly reduced, which enables us to better constrain the water ice column densities than those estimated from our previous \textit{AKARI} observations.

\item
The 3.53 $\mu$m CH$_3$OH ice absorption band for the LMC YSOs is found to be absent or weak compared to those seen toward Galactic counterparts. 
The absorption band is marginally detected for two out of eleven objects. 
We estimate the abundance of CH$_3$OH ice relative to water ice, which suggests that solid CH$_3$OH is less abundant in the LMC high-mass YSOs than in Galactic sources. 

\item 
We propose that warm ice chemistry in the LMC is responsible for the low abundance of solid CH$_3$OH presented in this work as well as the relatively high abundance of solid CO$_2$ reported in previous observations \citep[e.g., ][]{ST10,Oli11}. 
When the dust temperature in a molecular cloud is high due to strong interstellar radiation field, hydrogenation of CO on the grain surface to form CH$_3$OH would become less efficient because of the decrease in available hydrogen atoms on the surface. 
On the other hand, the formation of CO$_2$ through the $CO + OH \to CO_2 + H$ reaction could be enhanced as a result of the increased mobility of parent species. 
This suppresses the CH$_3$OH production, whereas enhances the CO$_2$ production. 
Since dust temperature of molecular clouds is believed to be related to metallicity of the host galaxy, such warm ice chemistry is presumably one of the important characteristics of interstellar and circumstellar chemistry in low metallicity galaxies. 

\item 
The 3.47 $\mu$m absorption band, which is generally seen in embedded sources, is detected toward six out of eleven objects in the LMC. 
The 3.47 $\mu$m band and the H$_2$O ice band correlate similarly between the LMC and Galactic samples, but the LMC sources seem to require a slightly higher H$_2$O ice threshold for the presence of the 3.47 $\mu$m band. 
The LMC sources with small H$_2$O ice optical depths show significantly lower $\tau_{3.47}$/$\tau_{3.05}$ ratio than those of Galactic sources. 
The difference in the threshold may suggest that more shielded environment is necessary for the formation of the 3.47 $\mu$m band carrier in the LMC. 
For the LMC sources with relatively large H$_2$O ice optical depths, we found that the $\tau_{3.47}$/$\tau_{3.05}$ ratio is only marginally low compared to Galactic sources. 
This would suggest that, in well-shielded regions, the lower metallicity as well as the different elemental abundances and different interstellar environment of the LMC have little effect on the abundance ratio of the 3.47 $\mu$m band carrier and water ice. 

\item 
The formation of complex organic molecules in low metallicity environment of the LMC is discussed based on the above results. 
CH$_3$OH is believed to be a starting point for formation of more complex carbonaceous molecules. 
The low CH$_3$OH ice abundance in the LMC implies that formation of complex organic molecules from methanol-derived species is less efficient in the LMC. 
However, formation processes of complex organic molecules in space are not well understood, and thus further observational, theoretical, and laboratory studies are needed to constrain the complex molecular chemistry in the past metal-poor universe. 

\end{enumerate}

We stress that the above conclusions are based on observations of a small number of extragalactic YSO samples. 
Further spectroscopic observations with a larger number of samples with higher sensitivity and higher spectral resolution are critically needed for comprehensive understanding of ice chemistry in low metallicity environments.

\begin{acknowledgements}
This research is based on observations with the Very Large Telescope at the European Southern Observatory, Paranal, Chile (programme ID 090.C-0497). 
We thank the VLT and ISAAC operation team members. 
We are especially grateful to the support astronomers for many helpful comments on site during our observations. 
This work uses data obtained by the \textit{AKARI} satellite, a JAXA project with the participation of ESA, and those obtained by NASA's \textit{Spitzer} Space Telescope. 
We are grateful to all the members who contributed to these projects.
This work has made extensive use of the Leiden Ice Database. 
This work is supported by the Japan Society for the Promotion of Science (JSPS) Research Fellow (24006664) and a Grant-in-Aid from the JSPS (15K17612). 
Finally, we would like to thank our referee, Jacco Th. van Loon, whose suggestions greatly improved this paper. 
\end{acknowledgements}

%-------------------------------------------------------------------

\appendix
\section{Comparison of \textit{AKARI} and VLT spectra} 
%%%%%%%%%%
\begin{figure*}[!]
\begin{center}
\includegraphics[width=13.7cm, angle=0]{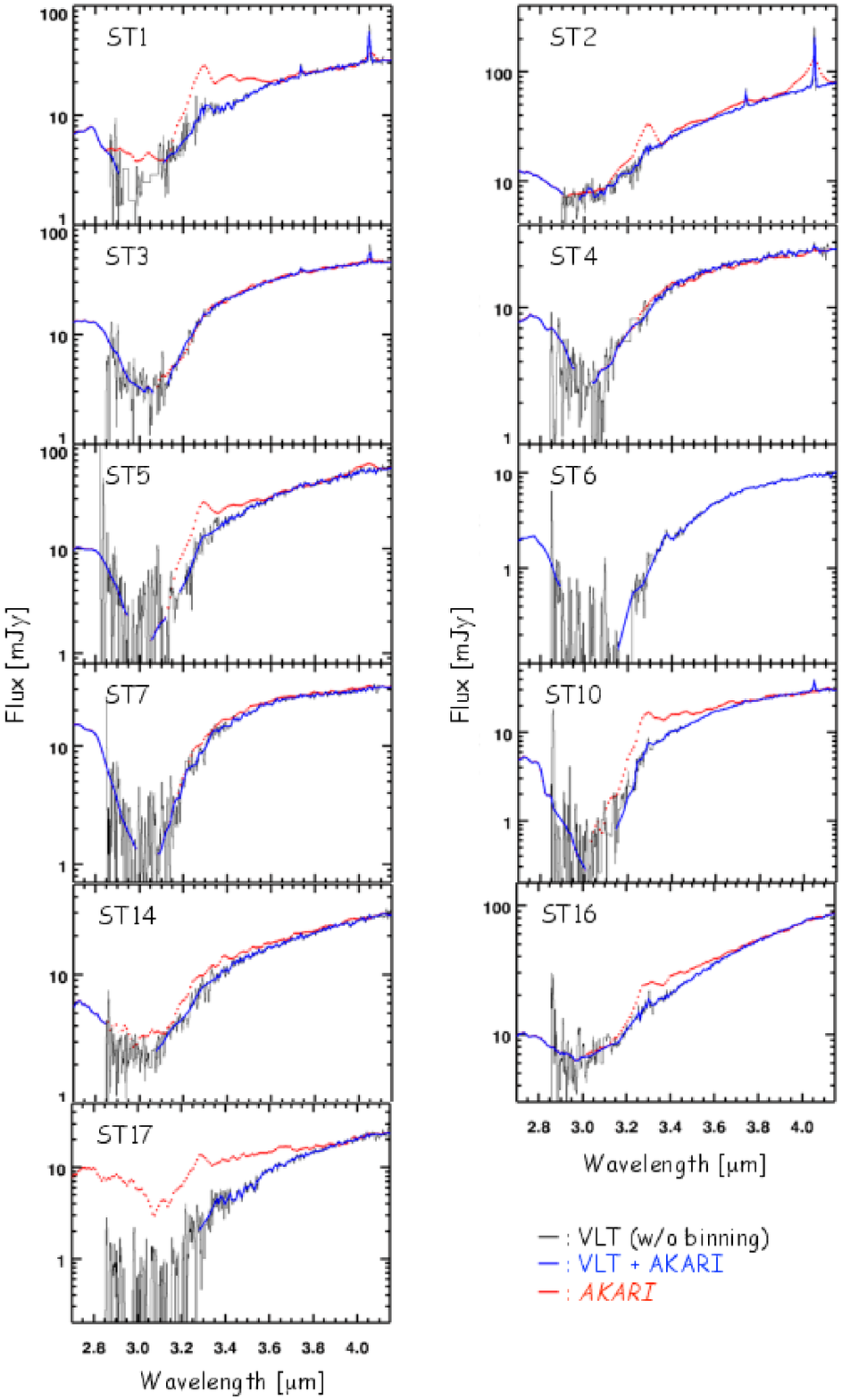}
\caption{Comparison of VLT/ISAAC and \textit{AKARI}/IRC spectra. 
The thin solid lines (black) represent the VLT spectra before applying spectral binning. 
The thick lines (blue) represent concatenated VLT and \textit{AKARI} spectra after binning. 
The circles (red) represent the \textit{AKARI} spectra before concatenation with VLT spectra. 
A detailed description about the spectra is given in $\S$3.1. 
A color version of this figure is available in the online journal.
}
\label{App_Spec}
\end{center}
\end{figure*}
%%%%%%%%%%

\section{Spectral fitting of CO$_2$ and CO ices} 
%%%%%%%%%%
\begin{figure*}[h]
\begin{center}
\includegraphics[width=13.0cm, angle=0]{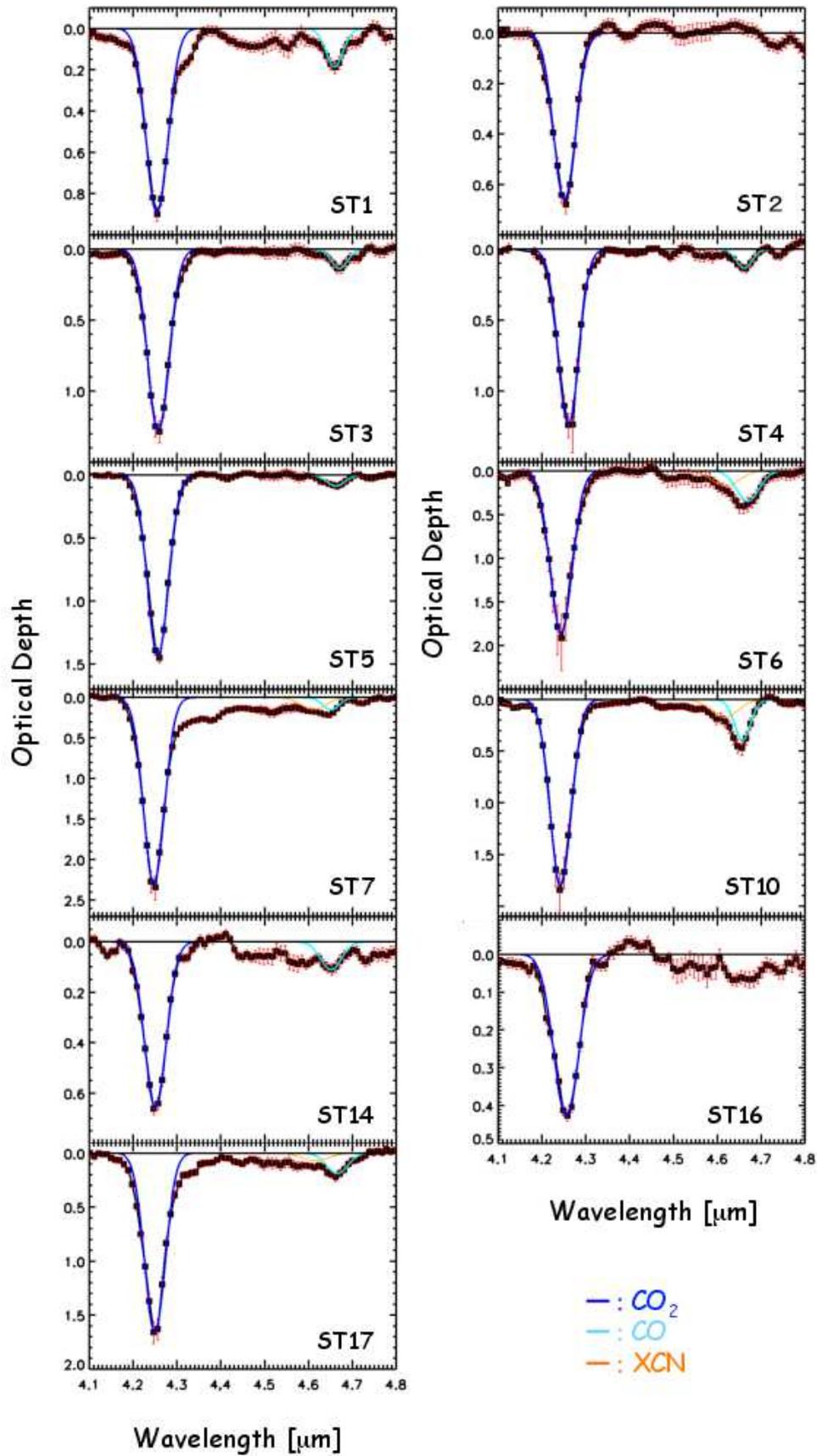}
\caption{Results of spectral fitting for the CO$_2$ and CO ice absorption bands of high-mass YSOs in the LMC. 
The blue solid lines represent the 4.27 $\mu$m CO$_2$ ice absorption band. 
The cyan solid lines represent the 4.67 $\mu$m CO ice absorption band. 
The orange thin solid lines represent the 4.62 $\mu$m XCN absorption band. 
Details of the fitting method is presented in \citet{ST10}. 
A color version of this figure is available in the online journal.
}
\label{App_Spec2}
\end{center}
\end{figure*}
%%%%%%%%%%

\end{document}